%% file: root.tex
\documentclass[journal]{IEEEtran}
\ifCLASSINFOpdf
\else
\fi
\usepackage{hyperref}
\usepackage{bbold}
\usepackage{booktabs}
\usepackage{amsmath} % assumes amsmath package installed
\usepackage{amssymb}  % assumes amsmath package installed
\usepackage{subcaption}
\newtheorem{proposition}{Proposition}
\usepackage{mathtools}
\usepackage{comment}
\usepackage{pgfplots}
\usepackage{xcolor}
\newtheorem{remark}{Remark}
\pgfplotsset{compat=newest}
\usetikzlibrary{plotmarks}
\usetikzlibrary{arrows.meta}
\usepgfplotslibrary{patchplots}
\usepackage{grffile}
\usepackage{tikz}
\usetikzlibrary{automata, positioning, arrows}
\usetikzlibrary{shapes,decorations}
\usetikzlibrary{positioning}
\usepackage{array}
\usepackage{breakcites}
\usepackage{float}
\newcommand{\added}{\textcolor{black}}
\newcommand{\addedTwo}{\textcolor{black}}
\newcommand{\addedThree}{\textcolor{black}}

\tikzset{node distance=2cm, % Minimum distance between two nodes. Change if necessary.
         every state/.style={ % Sets the properties for each state
            ellipse,
            semithick,
            fill=gray!10,
            minimum height=2cm,
            minimum width=2.75cm},
         initial text={},     % No label on start arrow
         double distance=4pt, % Adjust appearance of accept states
         every edge/.style={  % Sets the properties for each transition
            draw,
            ->,>=stealth',     % Makes edges directed with bold arrowheads
            auto,
            semithick}}

\newcommand{\R}{{\mathbb R}}

\begin{document}
%
% paper title
% Titles are generally capitalized except for words such as a, an, and, as,
% at, but, by, for, in, nor, of, on, or, the, to and up, which are usually
% not capitalized unless they are the first or last word of the title.
% Linebreaks \\ can be used within to get better formatting as desired.
% Do not put math or special symbols in the title.
\title{Improving Urban Traffic Throughput with Vehicle Platooning: Theory and Experiments}
%
%
% author names and IEEE memberships
% note positions of commas and nonbreaking spaces ( ~ ) LaTeX will not break
% a structure at a ~ so this keeps an author's name from being broken across
% two lines.
% use \thanks{} to gain access to the first footnote area
% a separate \thanks must be used for each paragraph as LaTeX2e's \thanks
% was not built to handle multiple paragraphs
%

\author{Stanley W. Smith$^{*,1}$,
        Yeojun Kim$^{*,2,3}$,
        Jacopo Guanetti$^3$,
        Ruolin Li$^2$,
        Roya Firoozi$^2$,
        Bruce Wootton$^3$, \\
        Alexander A. Kurzhanskiy$^4$,
        Francesco Borrelli$^{2,3}$,
        Roberto Horowitz$^2$,
        and Murat Arcak$^{1,2}$
        % <-this % stops a space
        \thanks{$^*$The first two authors contributed equally to this work. This project was supported in part by the Department of Defense through the NDSEG Fellowship program and the National Science Foundation through grant CNS-1545116, co-sponsored by the Department of Transportation. The material in this paper was partially presented at the 2019 European Control Conference, June 25 - 28, Naples, Italy. Corresponding author S. W. Smith.}
        \thanks{$^{1}$Dept. of Electrical Engineering and Computer Sciences, University of California, Berkeley
        {\tt\footnotesize $\{$swsmth,arcak$\}$@eecs.berkeley.edu.}}%
        \thanks{$^{2}$Dept. of Mechanical Engineering, University of California, Berkeley
        {\tt\footnotesize $\{$yk4938,ruolin\_li,royafiroozi,horowitz,fborrelli$\}$ @berkeley.edu.}}
        \thanks{$^{3}$AV-Connect Inc
        {\tt\footnotesize $\{$jacopoguanetti,wootton$\}$@avconnect.ai.}}
        \thanks{$^{4}$California Partners for Advanced Transportation Technology, University of California, Berkeley
        {\tt\footnotesize akurzhan@berkeley.edu.}}%
}

\IEEEoverridecommandlockouts
\IEEEpubid{\makebox[\columnwidth]{
This work is licensed under a \href{http://creativecommons.org/licenses/by/4.0/}{Creative Commons Attribution 4.0 License.}
\hfill} \hspace{\columnsep}\makebox[\columnwidth]{ }}

% make the title area
\maketitle

% As a general rule, do not put math, special symbols or citations
% in the abstract or keywords.

\begin{abstract}
In this paper we present a model-predictive control (MPC) based approach for vehicle platooning in an urban traffic setting. Our primary goal is to demonstrate that vehicle platooning has the potential to significantly increase throughput at intersections, which can create bottlenecks in the traffic flow. To do so, our approach relies on vehicle connectivity: vehicle-to-vehicle (V2V) and vehicle-to-infrastructure (V2I) communication. In particular, we introduce a customized V2V message set which features a velocity forecast, i.e. a prediction on the future velocity trajectory, which enables platooning vehicles to accurately maintain short following distances, thereby increasing throughput. Furthermore, V2I communication allows platoons to react immediately to changes in the state of nearby traffic lights, e.g. when the traffic phase becomes green, enabling additional gains in traffic efficiency. We present our design of the vehicle platooning system, and then evaluate performance by estimating the potential gains in terms of throughput using our results from simulation, as well as experiments conducted with real test vehicles on a closed track. Lastly, we briefly overview our demonstration of vehicle platooning on public roadways in Arcadia, CA.
\end{abstract}


\maketitle

\section{Introduction}
Vehicle connectivity and autonomy are important areas of research, both of which have made a notable impact on the automotive industry \cite{guanetti2018control}. For example, advanced driver assist systems (ADAS) which automate the longitudinal and lateral motion of the vehicle, such as the Tesla Autopilot and Cadillac Super Cruise systems, are being offered as an option in an increasing number of production vehicles. Furthermore, V2V communication technology is now included as a standard feature in Cadillac CTS sedans \cite{cadillacV2Varticle}.

The advent of connected automated vehicles has also paved the way towards significant improvements in transportation broadly \cite{uhlemann2016platooning}, including increased safety (by allowing, for example, the detection of vehicles occluded from sight) and reduced reliance on traffic lights \cite{V2Varticle}. V2V communication allows for nearby vehicles to coordinate their motion accurately and to form \textit{vehicle platoons}: strings of vehicles driving at the same speed and at short distance. There are two primary benefits of vehicle platooning: an improvement in traffic efficiency due to increased roadway capacity, and an increase in fuel efficiency due to reduced aerodynamic drag forces acting on the platooning vehicles, especially for heavy-duty vehicles such as semi-trucks. Regarding the first point, there is demonstrated potential for platooning to increase the capacity of both highways and urban roadways. For example, a microscopic simulation study in \cite{shladover2012impacts} predicts that increasing the penetration of vehicles capable of cooperative adaptive cruise control (CACC) will result in an increase in highway capacity, since it enables the driver to select smaller time headways. In \cite{lioris2017platoons} the authors predict that the throughput of urban roadways could potentially be doubled by forming platoons of vehicles, particularly by increasing the capacity of intersections, which they confirm with a subsequent simulation study. For the second point, experiments presented in \cite{bonnet2000fuel} confirm that small spacings between two heavy-duty trucks results in reduced fuel consumption.

\begin{figure}
    \centering
    \includegraphics[width = \columnwidth]{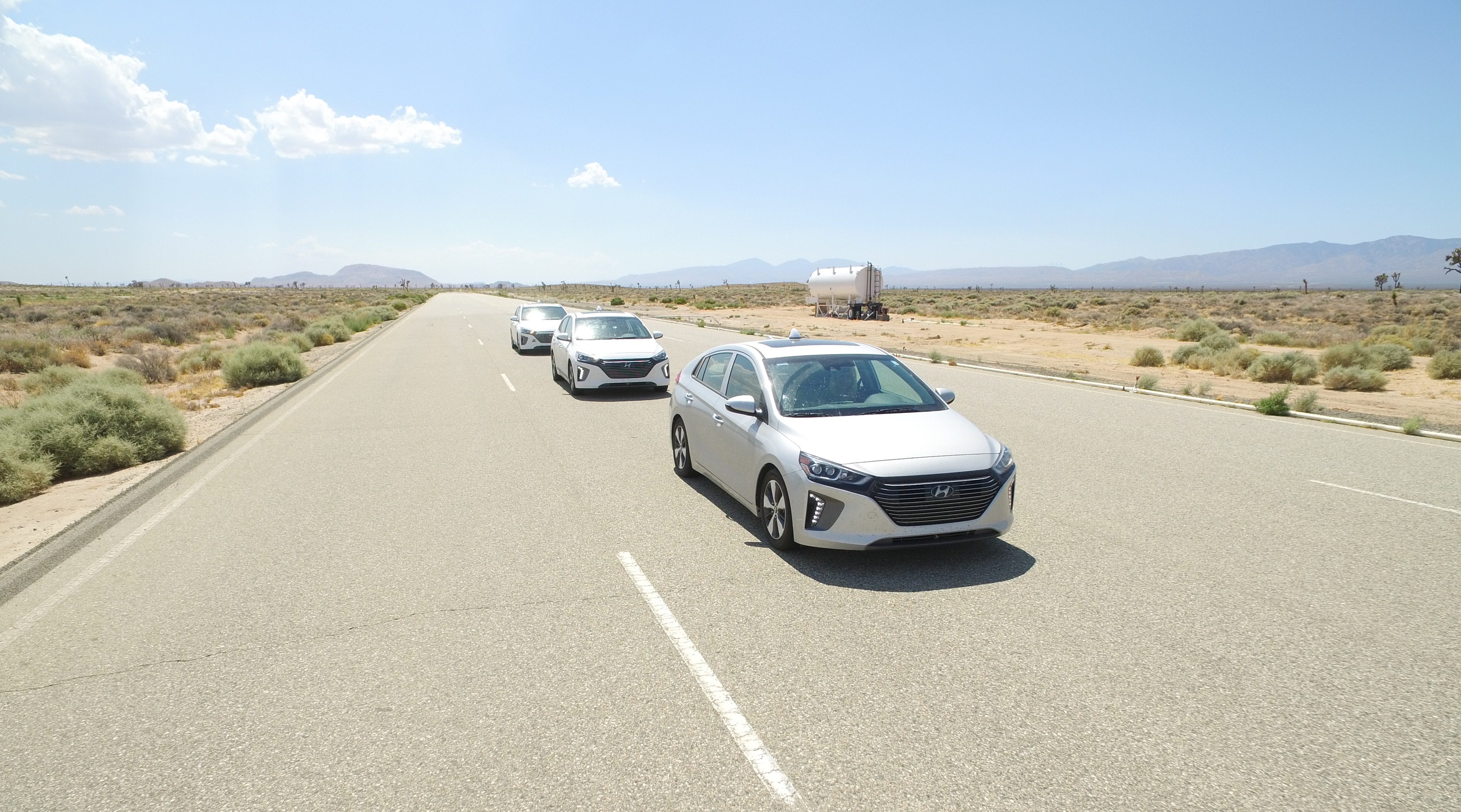}
    \caption{Test vehicles at the Hyundai-KIA Motors California Proving Grounds, California City, CA.}
    \label{fig:ioniqs}
\end{figure}

Previous demonstrations have showcased the technical feasibility of vehicle platooning. For example, vehicle platooning was demonstrated in 1994 and 1997 by the California PATH team on the I-18 highway in San Diego, CA \cite{shladover2007path}. Other experimental evaluations conducted on highways include \cite{alam2015heavy}, where the authors develop a platooning system architecture for heavy-duty vehicles. The system is evaluated in terms of controller tracking performance and fuel consumption over varying levels of road grade. In \cite{ploeg2011design, milanes2013cooperative} the authors present the design of a CACC system and tested it on a fleet of test vehicles. A primary controller performance metric in these works is \textit{string stability} \cite{swaroop1997string}, meaning that the preceding vehicles are able to attenuate disturbances in traffic downstream (for example, changes in velocity). In 2011 the first Grand Cooperative Driving Challenge was held in the Netherlands \cite{ploeg2012introduction}, with the goal of accelerating the deployment of cooperative driving technologies. The competition focused on CACC and included both an urban and highway driving challenge \cite{van2012cooperative}. For the urban driving challenge one criterion used to judge the participating teams was throughput improvement at the traffic light. This scenario is similar to the one we considered in our previous work \cite{smith2019balancing}, where we focused on the trade-off between traffic throughput gains and safety.

In addition to maintaining a platoon formation, the related tasks of forming, merging, and splitting platoons require structured coordination between vehicles, i.e. interaction protocols, which can be achieved in principle with V2V communication. For example, in \cite{hsu1991design} state machines are provided which describe the sequence of events, coordinated via V2V communication, that must occur during merge, split, and change lane maneuvers. Furthermore, low level control laws for the leader vehicle to execute these maneuvers have been developed \cite{li1997ahs}. In \cite{van2015d3} an extended message set is proposed for the purpose of enabling connected vehicles to coordinate more complex maneuvers in merging, intersection, and emergency vehicle scenarios for a follow-up Grand Cooperative Driving Challenge which was held in 2016 \cite{ploeg2018guest}. Other works studying communication include \cite{dolk2017event}, where the authors present a strategy for maintaining string stability in a vehicle platoon while using significantly fewer communication resources.

Unlike the aforementioned studies, in this work we focus on advancing vehicle platooning to a public urban environment where increased intersection throughput can result in significant improvements in overall traffic efficiency. Enabling platooning in an urban environment involves addressing various challenges, such as forming and disbanding platoons in moving traffic, decision-making (e.g. whether or not to proceed through an upcoming intersection), and ensuring safety when a lead vehicle is present. \addedTwo{These challenges are especially important on a public roadway, where the future behavior of vehicles ahead of the platoon and the phase of upcoming traffic lights are uncertain}. We present a design for the urban platooning system, and then analyze performance by estimating throughput using data obtained from simulations and experiments conducted on a closed track. \addedTwo{We also introduce a state machine for managing the participating platooning vehicles, and propose strategies for the platoon to ensure safety when it encounters an intersection and / or a leading vehicle, utilizing predictions of their future behavior}.

The closest comparable effort that we are aware of is the MAVEN project, which has laid out the various technologies that are needed to develop and deploy urban platooning, and reported on test results with two automated vehicles \cite{schindler2020maven}, where technologies such as a green light optimal speed advisory system and a collective perception message were utilized. \addedTwo{Unlike \cite{schindler2020maven}, our focus in this paper is on improving throughput by maintaining short (constant) distances between the vehicles as the platoon accelerates from rest to a nominal speed. In particular, we achieve such accurate tracking by transmitting velocity forecasts between platooning vehicles and using them as disturbance previews in our MPC problems}.

\added{The remainder of the paper is organized as follows. We outline our design for the urban vehicle platooning system in Sections \ref{vehicles} - \ref{safetyUrban}, including a platoon model \addedTwo{and management system}, MPC formulation, and strategy for the leader to ensure safety. In Section \ref{simulation} we present results from our simulation tool, and analyze the performance of the platooning system by estimating the potential gains in intersection throughput. Next, in Section \ref{experimentalResults} we discuss the experimental setup and present results from conducting tests on a closed track and on public roadways in Arcadia, CA, including our estimates of throughput. We end with concluding remarks in Section \ref{conclusion} and discuss some of the challenges we encountered during the tests, as well as potential solutions. We note that parts of Sections \ref{vehicles} - \ref{safetyUrban} are adapted from our previous work \cite{smith2019balancing}, but the remaining content in the paper is completely new and advances platooning to an urban setting}.

\section{Platoon Model and Management} \label{vehicles}
\addedTwo{In this section we introduce the model of the platoon and various systems that enable management of its behavior (beyond the control algorithms themselves), including state estimation via on-board sensors, V2X communication, and a finite-state machine (FSM) system which ensures that the platoon acts in a coordinated manner, that is, vehicles start moving as a single platoon at the same time and break the platoon at the same time as needed. In particular, we discuss how vehicle-to-vehicle communication enables the follower vehicles to do accurate distance tracking of the leader, and how vehicle-to-infrastructure communication enables the leader to decide whether or not to proceed through an upcoming intersection}.

\subsection{Vehicle Models} \label{vehDyn}
The longitudinal dynamics of the leader vehicle \cite{guzzella2007vehicle} are modelled as
\begin{subequations} \label{leadDyn}
\begin{align}
\dot{p}_L(t) &= v_L(t), \\
\dot{h}_L(t) &= v_F(t) - v_L(t), \\
\dot{d}_L^{TL}(t) &= -v_L(t), \\
\dot{v}_L(t) &= \frac{1}{M} \left( \frac{T_L^a(t) - T_L^b(t)}{R_w} - F_f(t) \right), \label{accelLeader} \\
\dot{T}_L^a(t) &= \frac{1}{\tau} \left( T_L^{a,ref}(t) - T_L^a(t) \right), \label{delayLeader}
\end{align}
\end{subequations}
where the states are as follows: $p_L(t)$ is the position, $h_L(t)$ is the distance to the public vehicle ahead (specifically, the distance from the front bumper of the leader vehicle to the rear bumper of the front vehicle), $d_L^{TL}(t)$ is the distance to the nearest upcoming intersection stop bar, $v_L(t)$ is the ego vehicle velocity, and $T_L^a(t) \in \R_{\geq 0}$ is the accelerating wheel torque. The inputs $T_L^{a,ref}(t) \in \R_{\geq 0}$ and $T_L^b(t) \in \R_{\geq 0}$ are the accelerating wheel torque command and the braking wheel torque. Lastly, $v_F(t)$ is the velocity of the public vehicle ahead, henceforth referred to as the front vehicle. The parameters $M$, $R_w$, and $\tau$ are the vehicle mass, wheel radius, and actuation time constant for acceleration, respectively. We note that \eqref{delayLeader} models actuation delay while the vehicle is accelerating, which has been empirically estimated by collecting wheel torque measurements from the test vehicle. During these experiments we observed no delay while braking, and therefore the model does not include actuation delay while braking. Lastly, $F_L^f(t)$ is a longitudinal force acting on the leader vehicle, given by
\begin{equation} \label{forceFriction}
F_L^f(t) = M g \left( (\sin(\theta) + r \cos(\theta) \right) + \frac{1}{2} \rho A c_x v_L(t)^2
\end{equation}
where $g$ is the gravitational constant, $\theta$ is road grade, $A$ is the area of the vehicle, $r$ is a rolling coefficient of the vehicle, $\rho$ is air density, and $c_x$ is an air drag coefficient. We assume road grade is negligible, and thus $\theta = 0$ for $t \geq 0$. For simplicity, we represent \eqref{forceFriction} as
\begin{equation} \label{alphaBetaGammaModel}
F_L^f(t) = \beta + \gamma v_L(t)^2
\end{equation}
where the parameters $\beta$, $\gamma \in \R_{\geq 0}$ were identified by collecting driving data at a testing area near UC Berkeley, and then fitting predictions from \eqref{alphaBetaGammaModel} to the data (see Table \ref{modelParameters}). We write the leader vehicle dynamics \eqref{leadDyn} concisely as
\begin{equation}
\dot{x}_L(t) = f_L(x_L(t), u_L(t), w_L(t)) \label{leadEqn}
\end{equation}
where $x_L(t) := [p_L(t); \ h_L(t); \ d_L^{TL}(t); \ v_L(t); \ T_L^a(t)]$, $u_L(t) := [T_L^{a,ref}(t); \ T_L^b(t)]$, and $w_L(t) := v_F(t)$. Note that the velocity of the front vehicle $v_F(t)$ appears as a disturbance here. Since we cannot accurately predict the behavior of non-platooning vehicles, we make the conservative assumption that the front vehicle will decelerate from its current speed until coming to a stop. This assumed trajectory of the front vehicle is used for planning, to be discussed further in Section \ref{leadMPC}.

\begin{table}[t]
    \caption{Model Parameters}
    \label{modelParameters}
    \centering
    \begin{tabular}{c l l l}
    \toprule
         $M$ & vehicle mass & kg & 2044 \\
         $R_w$ & wheel radius & m & 0.3074 \\
         $\beta$ & frictional force modelling parameter & - & 339.1329 \\
         $\gamma$ & (same as above) & - & 0.77 \\
         $\tau$ & accelerating torque actuation time constant & s & 0.7868 \\
         $\Delta t$ & sampling time & s & 0.1 \\
    \bottomrule
    \end{tabular}
\end{table}

We model the longitudinal dynamics of each of the $N - 1$ follower vehicles in the platoon as
\begin{subequations} \label{followDyn}
\begin{align}
\dot{p}_i(t) &= v_i(t), \\
\dot{h}_i(t) &= v_{i-1}(t) - v_i(t), \\ \label{followerHeadway}
\dot{s}_i(t) &= v_L(t) - v_i(t), \\
\dot{v}_i(t) &= \frac{1}{M} \left( \frac{T_i^a(t) - T_i^b(t)}{R_w} - F_f(t) \right), \label{accelFollower} \\
\dot{T}_i^a(t) &= \frac{1}{\tau} \left( T_i^{a,ref}(t) - T_i^a(t) \right), \quad i = 1, \dots, N, \label{delayFollower}
\end{align}
\end{subequations}
where $s_i(t)$, used for distance tracking relative to the leader vehicle, is defined as follows:
\begin{equation}
s_i(t) = \sum_{k = 1}^i h_k(t). \label{leadDist}
\end{equation}
We refer to $s_i(t)$ as the distance from follower $i$ to the leader (note that \eqref{leadDist} implies $s_1(t) = h_1(t)$). Furthermore, we let $v_0(t) = v_L(t)$ so that \eqref{followerHeadway} is valid for follower $i = 1$. We write \eqref{followDyn} compactly as
\begin{equation}
\dot{x}_i(t) = f_i(x_i(t), u_i(t), w_i(t)), \quad i = 1, \dots, N - 1, \label{followerEqn}
\end{equation}
where $x_i(t) := [p_i(t); \ h_i(t); \ d_i^{TL}(t); \ v_i(t); \ T_i^a(t)]$, $u_i(t) := [T_i^{a,ref}(t); \ T_i^b(t)]$, and $w_i(t) := [v_L(t); \ v_{i-1}(t)]$. We note that the velocity of the leader and front vehicle $v_L(t)$ and $v_{i-1}(t)$ appear as disturbances here  - since these are both platooning vehicles in this case, we can receive a forecast of their future behavior via V2V communication. In Section \ref{v2vcomm} we discuss the information transmitted between platooning vehicles which includes a velocity forecast, to be used as a disturbance preview in our MPC formulation.

For planning, our goal is to obtain linear, discrete time models from \eqref{leadEqn} and \eqref{followerEqn}. We use the procedure outlined in \cite{smith2019balancing} for doing so: we first linearize the leader and follower vehicle dynamics about the nominal velocities $v_L^0$ and $v_i^0$, respectively, and then discretize the resulting linear models each with time step $\Delta t = 0.1$s, resulting in
\begin{align}
x_L(k + 1) &= A_L x_L(k) + B_L u_L(k) + E_L w_L(k), \nonumber \\[2.5pt]
x_i(k + 1) &= A_i x_i(k) + B_i u_i(k) + E_i w_i(k),
\end{align}
where the matrices $A_L \in \R^{5 \times 5}$ and $B_L \in \R^{5 \times 2}$ are functions of the velocity $v_L^0$, and $A_i \in \R^{6 \times 6}$ and $B_i \in \R^{6 \times 2}$ are functions of the velocity $v_i^0$. At each time step, the current ego vehicle velocity is substituted into these expressions to obtain the appropriate dynamics matrices to be used for MPC.

\begin{figure}
\centering
\includegraphics[width = \columnwidth]{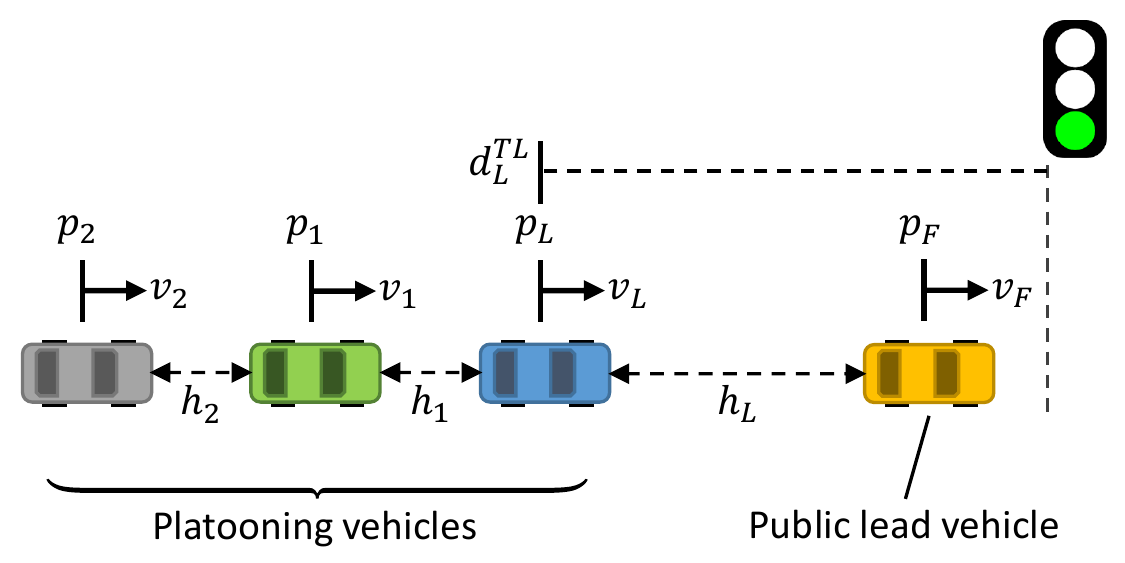}
\caption{Depiction of the states for a platoon of size $N = 3$ and public lead vehicle approaching an upcoming traffic light.} \label{fig:platoonDiagram}
\end{figure}

\subsection{State Estimation} \label{stateEstimation}
To localize the leader and follower vehicle positions $p_L(t)$ and $p_i(t)$ we use a differential GPS measurement which has lane-level accuracy. Furthermore, with GPS and information received from nearby traffic lights we can also estimate the distances $d_L^{TL}(t)$ and $d_i^{TL}(t)$ from each vehicle to the nearest upcoming traffic light. The forward-looking radar on each vehicle measures the headways $h_L(t)$ and $h_i(t)$, and standard on-board sensors provide the current velocity estimates $v_L(t)$ and $v_i(t)$, as well as estimates of the accelerating wheel torques $T_L^a(t)$ and $T_i^a(t)$.

An important sensing challenge for each follower $i$ is to estimate the distance to the leader as defined in \eqref{leadDist}. We have tested two methods for doing so: 1) estimating $s_i(t)$ using GPS, and 2) estimating $s_i(t)$ directly using the radar measurements $h_i(t)$, which can be transmitted via V2V communication. For the first method, we use GPS to measure the distance $d_i^L(t)$ from the center of vehicle $i$ to the center of the leader vehicle and use the estimate
\begin{equation} \label{method1}
\hat{s}_i(t) = \hat{d}_i^L(t) - i \cdot L_{\text{veh}}
\end{equation}
where $\hat{d}_i^L(t)$ is an estimate of $d_i^L(t)$ from GPS. The main drawback to this approach is GPS measurement noise - we observed up to 3 meters of error when estimating $s_i(t)$ using GPS. Because of this, we also used a Kalman filter, where the idea is to use the current velocity of the leader (received via V2V communication) and the ego vehicle velocity to improve our estimate of $s_i(t)$. For the second method, we use the estimate
\begin{equation} \label{method2}
\hat{s}_i(t) = \sum_{k = 1}^i \hat{h}_k(t),
\end{equation}
where $\hat{h}_k(t)$ is an estimate of $h_k(t)$ from radar. Since measurements from the forward-looking radar are generally very reliable, we observed smaller measurement errors using the second method. The main drawback to the second approach, however, is that it will require more vehicles in the platoon to communicate with one another (discussed further in the next section). For the experiments discussed in Section \ref{CPG} we used GPS to estimate $s_i(t)$, and for the experiments in Section \ref{arcadia} we used radar measurements to estimate $s_i(t)$.

\subsection{Vehicle-to-vehicle communication} \label{v2vcomm}
\addedTwo{We assume each platooning vehicle is capable of} V2V communication. An important piece of information transmitted within the platoon is a forecast of the future velocity trajectory for each vehicle, given by
\begin{align}
v_L^{\text{forecast}} &= [v_L(t|t); \ v_L(t+1|t); \ \dots; \ v_L(t+N_p|t)], \nonumber \\
v_i^{\text{forecast}} &= [v_i(t|t); \ v_i(t+1|t); \ \dots; \ v_i(t+N_p|t)], \label{velForecasts}
\end{align}
for the leader vehicle and follower vehicle $i$, respectively. Here, $v_L(k|t)$ is the planned velocity of the leader vehicle at time step $k$, obtained by solving an MPC problem at the current time step $t$ (the notation is the same for the follower vehicles), and $N_p$ is the MPC horizon in time steps. Each follower vehicle receives a velocity forecast from the front vehicle and the leader vehicle, corresponding to the flow of information depicted in Figure 3a. The front vehicle forecast is used to ensure safety, and the leader vehicle forecast is used to do distance tracking of the leader.

In addition to the velocity forecast, each experimental vehicle transmits its radar measurement, current GPS coordinates, and plan status signal. A secondary reason for transmitting GPS coordinates, beyond estimating $s_i(t)$, is so that the leader vehicle can estimate the distance $d_L^{N-1}(t)$ from itself to the rear platooning vehicle. The transmission of GPS coordinates from follower $N - 1$ to the leader is shown in Figure 3b. This lets the leader check whether the entire platoon has enough time to pass through an upcoming intersection, as discussed in the next section. As mentioned in the previous section, for some of our experiments we used radar measurements, transmitted via V2V communication, to estimate $s_i(t)$. In Figure 3c we depict the flow of information in this case, for $N = 4$. \added{We note that each vehicle, upon receiving an incoming message, checks the ID of the vehicle that transmitted it (indicating the vehicle's position in the platoon, e.g. leader vehicle, rear vehicle, etc.) to determine which information fields to extract, if any}.

\begin{figure}
\centering
\includegraphics[width = 0.85\columnwidth]{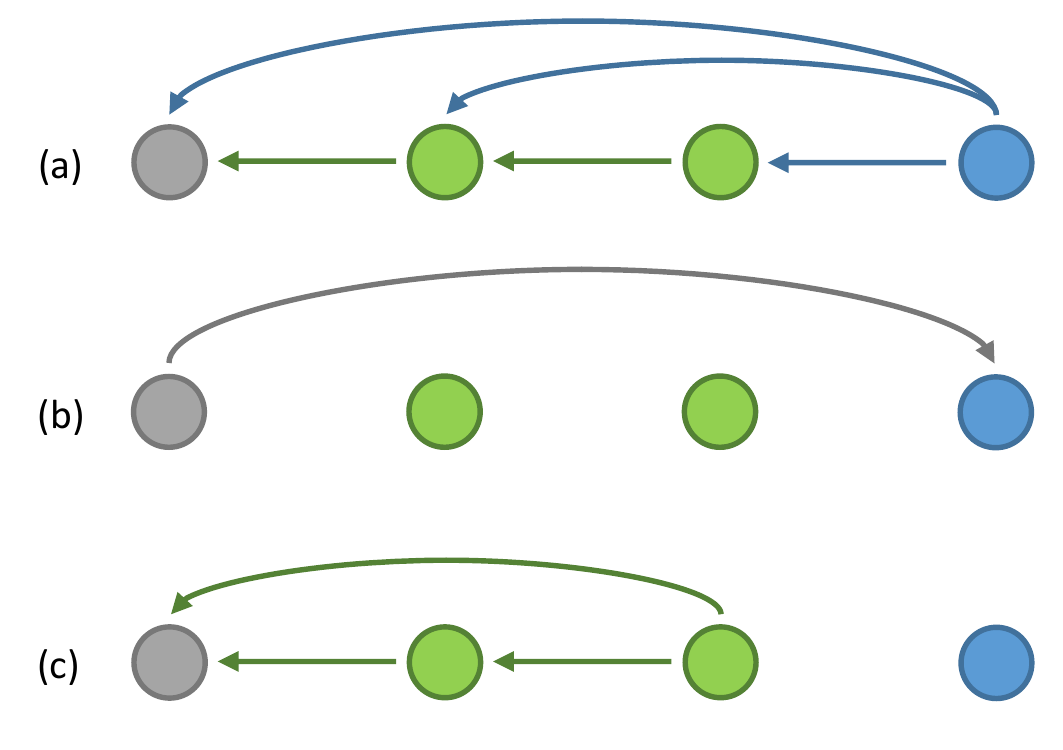}
\caption{Flow of V2V messages for a platoon of size $N = 4$, where the blue node represents the leader vehicle and the grey node represents the rear vehicle. Figure 3a shows the transmission of velocity forecasts and 3b shows the transmission of GPS coordinates from the rear vehicle (used by the leader to determine if the platoon can make it through the intersection, see Section \ref{v2icomm}). Figure 3c shows how we share radar measurements when we use the second method for estimating $s_i(t)$ as in \eqref{method2}.} \label{fig:v2vFlow}
\end{figure}

\subsection{Vehicle-to-infrastructure communication} \label{v2icomm}
In addition to V2V messages, \addedTwo{we assume the platooning vehicles} also receive SPaT (signal, phase, and timing) messages from nearby traffic lights via V2I communication. \addedTwo{In this way, each vehicle obtains the following prediction on the nearest upcoming traffic light state}:
\begin{equation}
\hat{x}_{TL}(t) = [p_{\text{up}}(t); \ c_r(t)]
\end{equation}
where $p_{\text{up}}(t) \in \{\text{red, yellow, green}\}$ is the current phase of the nearest upcoming traffic light and $c_r(t) \in \R_{\geq 0}$ is a prediction on the time remaining in the current phase. We note that it is necessary to predict $c_r(t)$ here since in our experiments the traffic signals are actuated.

\addedTwo{In the remainder of this section, we discuss how the leader decides whether or not the platoon should stop at an upcoming traffic light. This decision is handled by the leader only - the follower vehicles simply track the leader, and therefore we do not allow platoon separation}. Suppose the platoon is approaching a traffic light during its green phase, with $c_r(t)$ seconds remaining in the phase. In this scenario, the leader checks if the following condition holds
\begin{equation} \label{leadShouldStop}
c_r(t) \cdot v_L(t) \geq d_L^{N-1}(t) + d_L^{TL}(t) + L_{int} 
\end{equation}
to determine whether a stop is necessary (specifically, if \eqref{leadShouldStop} is false the platoon should stop), where $L_{\text{int}}$ is the intersection length. Condition \eqref{leadShouldStop} provides a quick and simple way to check whether the rear platooning vehicle, travelling at the current leader velocity $v_L(t)$, will pass through the intersection during the green phase. We use $v_L(t)$ in \eqref{leadShouldStop} since the leader effectively sets the speed for all platooning vehicles behind it, and also to avoid having to transmit $v_{N-1}(t)$ to the leader. We note that when $N$ is large, $d_L^{N-1}(t)$ is large and thus \eqref{leadShouldStop} is easily violated. This means the platoon may begin braking during a green light, which can be unexpected for nearby drivers. To avoid this, for large $N$ allowing platoon separation \addedTwo{may become necessary}.

\added{At low velocity \eqref{leadShouldStop} is not easily satisfied and will be overly restrictive, for example if the light just turned green and the platoon is stopped}. For this reason, if $v_L(t) \leq v_{\text{low}}$ the leader simply checks if the following condition holds
\begin{equation} \label{leadShouldStopLow}
c_r(t) \geq t_{\min}
\end{equation}
where the threshold $t_{\min}$ is a tuning parameter. If so, it is considered safe to proceed. \addedTwo{By checking \eqref{leadShouldStop} and \eqref{leadShouldStopLow} to determine whether to stop, we try to ensure the platoon will not be crossing the intersection when the phase becomes yellow. However, since the traffic signal is actuated and can change randomly due to uncertain traffic conditions, we cannot formally guarantee that this will never occur}.

\addedTwo{Suppose the leader determines it should stop while the phase is green, or that the phase is yellow, in which case the leader should stop if it can do so safely}. Then, we also check if the leader is capable of stopping before the intersection stop line with a margin of $d_{\min}$, that is
\begin{equation} \label{leadCanStop}
\frac{v_L(t)^2}{2 a_\text{min,brake}} \leq d_L^{TL}(t) - d_{\min}
\end{equation}
where $-a_\text{min,brake} \in \R_{< 0}$ is \addedTwo{an upper bound on \eqref{accelLeader} while the maximum braking force is applied}. \added{If \eqref{leadCanStop} does not hold, then it is deemed safer for the leader to proceed through the intersection (in this scenario, for large $N$ a platoon separation may also be necessary). For a red phase, however, we require the platoon to stop in any case}.

\subsection{Finite state machine (FSM)} \label{statemanager}
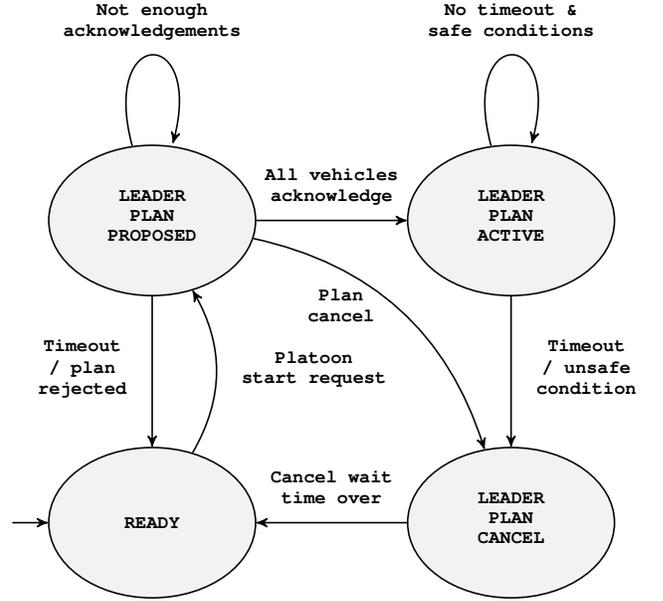
\begin{figure}

    \centering

    \begin{tikzpicture}
    
    \setlength\extrarowheight{-6pt}
    
    \node[state, ellipse, initial] (q1) {\tt\scriptsize \textbf{READY}};
                                         
    \node[state, ellipse, above = of q1] (q2) {\begin{tabular}{c}
                                                {\tt\scriptsize \textbf{LEADER}} \\
                                                {\tt\scriptsize \textbf{PLAN}} \\
                                                {\tt\scriptsize \textbf{PROPOSED}} \\
                                                \end{tabular}};
                                                
    \node[state, ellipse, right = of q2] (q3) {\begin{tabular}{c}
                                                {\tt\scriptsize \textbf{LEADER}} \\
                                                {\tt\scriptsize \textbf{PLAN}} \\
                                                {\tt\scriptsize \textbf{ACTIVE}} \\
                                                \end{tabular}};
                                                
    \node[state, ellipse, below = of q3] (q4) {\begin{tabular}{c}
                                                {\tt\scriptsize \textbf{LEADER}} \\
                                                {\tt\scriptsize \textbf{PLAN}} \\
                                                {\tt\scriptsize \textbf{CANCEL}} \\
                                                \end{tabular}};
    
    \setlength\extrarowheight{-6pt}
    
    \draw   (q1) edge[bend right] node[right]{\begin{tabular}{c}
                                    {\tt\scriptsize \textbf{Platoon}} \\
                                    {\tt\scriptsize \textbf{start request}}
                                    \end{tabular}} (q2)
                                
            (q2) edge node[left]{\begin{tabular}{c}
                                {\tt\scriptsize \textbf{Timeout}} \\
                                {\tt\scriptsize \textbf{/ plan}} \\
                                {\tt\scriptsize \textbf{rejected}}
                                \end{tabular}} (q1)
            
            (q2) edge[loop above] node{\begin{tabular}{c}
                                        {\tt\scriptsize \textbf{Not enough}} \\
                                        {\tt\scriptsize \textbf{acknowledgements}}
                                        \end{tabular}} (q2)
            
            (q2) edge[below] node[above]{\begin{tabular}{c}
                                    {\tt\scriptsize \textbf{All vehicles}} \\
                                    {\tt\scriptsize \textbf{acknowledge}}
                                    \end{tabular}} (q3)
            
            (q2) edge[bend left] node[left]{\begin{tabular}{c}
                                            {\tt\scriptsize \textbf{Plan}} \\
                                            {\tt\scriptsize \textbf{cancel}}
                                            \end{tabular}} (q4)
            
            (q3) edge[loop above] node[above]{\begin{tabular}{c}
                                             {\tt\scriptsize \textbf{No timeout \&}} \\
                                             {\tt\scriptsize \textbf{safe conditions}}
                                             \end{tabular}} (q3)
            
            (q3) edge node[right]{\begin{tabular}{c}
                                    {\tt\scriptsize \textbf{Timeout}} \\
                                    {\tt\scriptsize \textbf{/ unsafe}} \\
                                    {\tt\scriptsize \textbf{condition}}
                                    \end{tabular}} (q4)
    
            (q4) edge node[above]{\begin{tabular}{c}
                                    {\tt\scriptsize \textbf{Cancel wait}} \\
                                    {\tt\scriptsize \textbf{time over}}
                                    \end{tabular}} (q1);
    \end{tikzpicture}
    \caption{A diagram of the transitions in our finite state machine, shown here for the leader vehicle for simplicity. \label{fig:FSM}}
\end{figure}

We have designed a FSM (see Figure \ref{fig:FSM}) which acts as a mechanism for safely forming and maintaining a platoon. There are four primary states in our FSM: `Ready', `Plan Proposed', `Plan Active', and `Plan Cancel'. 
Each platooning vehicle is initialized in the `Ready' state and communicates its state at all times.
The platoon formation process is initiated when the leader moves to the `Plan Proposed' state by proposing to the follower vehicles the `plan', including a plan ID, ordering of the vehicles in the platoon, desired gap / speed, etc. Note that the ordering of vehicles in the platoon refers to the list of vehicle IDs ordered from the leader to the last follower. As soon as the `plan' is received by the followers, the states of the followers transition to the `Plan Proposed' state.
In the `Plan Proposed' state, each vehicle acknowledges that the `plan' is valid by checking the on-board sensor data and communicated GPS data. For example, each vehicle can confirm that the driver agrees to join the platoon and that the proposed `Plan' is safe to follow. We also note that the leader can manually cancel the plan while in the `Plan Proposed' state, forcing a transition to the `Plan Cancel' state.

When the leader receives an acknowledgement from every vehicle in the `Plan', it moves to the `Plan Active' state while also informing the followers so that all vehicles move to the `Plan Active' state together. To ensure safety, while in the `Plan Active' state every vehicle in the platoon continuously monitors the surrounding conditions to decide if the `Plan' must stop. In our experiments, the conditions that cancel the plan include: 1) incorrect ordering of the vehicles, 2) message timeout, 3) any driver taps the gas / brake pedal, 4) front vehicle out of range (radar measurement too high), and 5) velocity upper / lower bound violated. Here, message timeout refers to when a particular message has not been received for a period of time longer than a specified threshold. When one of these conditions is detected by one vehicle, it informs the other vehicles in the platoon and they move together to the `Plan Cancel' state. 
After some threshold time, each vehicle transitions from the `Plan Cancel' state to the `Ready' state and the platoon can be restarted as needed.

\begin{figure}
    \centering
    \input{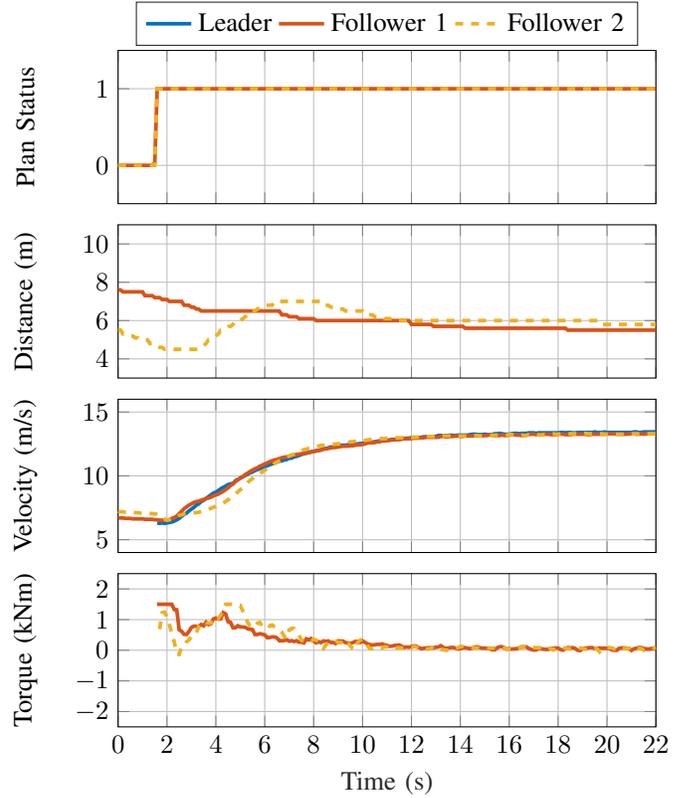}
    \caption{Experimental data collected in Arcadia, CA during the platoon formation process. The vehicles begin at a low speed and unequal spacing. At around the 2s mark, the platoon leader proposes a `plan' which is accepted by the following vehicles, and the plan status signal (plotted above) switches from 0 to 1. This engages all platooning controllers simultaneously, and the vehicles quickly converge to the desired speed and distance.}
    \label{fig:state_machine}
\end{figure}

In Figure \ref{fig:state_machine} we display some data collected while forming a platoon during testing in Arcadia, CA (see Section \ref{arcadia}). The procedure for forming a platoon was to manually drive the test vehicles to get them close together and moving at similar speeds, at which point the leader vehicle would propose a `plan' via the state machine and engage the platooning controllers simultaneously. This enabled platoon formation even while the vehicles are moving.

\section{MPC Formulation} \label{MPC}
In this section we present our MPC problem formulation for the platoon. The leader vehicle has a separate MPC problem which allows it to react to changing traffic conditions and set the desired velocity for the following vehicles. For example, if a stop at an intersection is necessary, the leader computes a velocity trajectory in order to stop safely and comfortably at the intersection stop bar. Furthermore, the leader maintains a safe following distance when a vehicle is present ahead of it. The follower vehicles simply do distance tracking relative to the leader, \addedTwo{as we do not allow platoon separation}.

\subsection{Leader vehicle MPC} \label{leadMPC}
The goal for the leader is to track a desired velocity when it is safe to do so. When necessary, it must yield to a slower-moving front vehicle or stop at the intersection stop bar. The MPC problem for the leader is
\begin{subequations} \label{leadMPCproblem}
\begin{align}
& \underset{u_i(\cdot|t)}{\text{min}}
& J_L & = \sum_{k=t}^{t+N_p+1} (v_L(k|t) - v_L^{des})^2 \label{leaderObj1} \\
& & & + \sum_{k=t}^{t+N_p} u_L(k|t)^T R u_L(k|t) \label{leaderObj2} \\
& & & + \alpha \sum_{k=t}^{t+N_p-1} \|u_L(k+1|t) - u_L(k|t)\|^2 \label{leaderObj3} \\
& \text{s.t.} & & x_L(k+1|t) = \label{leaderConstr1} \\
& & & \quad A_L x_L(k|t) + B_L u_L(k|t) + E_L \hat{w}_L(k), \nonumber \\
& & & v_{\min} \leq v_L(k|t) \leq v_{\max}, \label{leaderConstr2} \\
& & & d_{\min} + t_h v_L(k|t) \leq d_L^*(k|t), \label{leaderConstr3} \\
& & & 0 \leq T_L^{a}(k|t) \leq T_{\max}^a, \label{leaderConstr4} \\
& & & 0 \leq T_L^{a, ref}(k|t) \leq T_{\max}^a, \label{leaderConstr5} \\
& & & 0 \leq T_L^b(k|t) \leq T_{\max}^b, \label{leaderConstr6} \\
& & & x_L(t|t) = \hat{x}_L(t), \label{leaderConstr7} \\
& & & \forall k = t, \dots, t+N_p, \nonumber \\
& & & \begin{bmatrix}
d_L^*(t+N_p|t) \\ v_L(t+N_p|t)
\end{bmatrix} \in C(\hat{x}_L(t), \hat{v}_F(t), \hat{v}_F(t+N_p)), \label{leaderConstr8}
\end{align}
\end{subequations}
where $N_p$ is the MPC horizon in time steps, and $x_L(k|t)$ and $u_L(k|t)$ are the planned state and input of the leader vehicle at time step $k$, computed at time step $t$, respectively (the notation for the other states is the same). Furthermore, $d_L^*(k|t)$ is the distance from the leader vehicle to either the front vehicle or the upcoming intersection stop bar - whichever is a higher priority obstacle (the method for determining this is outlined in Section \ref{safetyUrban}). Lastly, $\hat{x}_L(t)$, $\hat{v}_F(t)$ are estimates of the leader vehicle and front vehicle state, based on measurements from the on-board sensors, and $\hat{v}_F(t + N_p)$ is an estimate of the front vehicle velocity at the end of the MPC planning horizon. Indeed, since $\hat{w}_L(k) := \hat{v}_F(k)$ appears as a disturbance in \eqref{leaderConstr1}, we must predict the future velocity trajectory of the front vehicle. To ensure safety, we assume worst-case behavior, i.e. the front vehicle will decelerate from its current speed at the rate $a_{\text{max,brake}} \in \R_{>0}$ until coming to a complete stop as follows
\begin{align}
& \hat{w}_L(k) := \hat{v}_F(k) = \label{frontBrakingMPC} \\
& \resizebox{\linewidth}{!}{ $ \quad \begin{cases}
\tilde{v}^0, & k = t, \\
\max(0, \ \hat{v}_F(k - 1) - k \cdot a_{\text{max,brake}} \cdot \Delta t), & k = t + 1, \dots, t + N_p,
\end{cases} $ } \nonumber 
\end{align}
where $\tilde{v}^0$ is an under-approximation of the front vehicle's current velocity $v^0$, to be discussed further in Section \ref{safetyUrban}. \addedTwo{Here, $-a_{\text{max,brake}} \in \R_{< 0}$ is a lower bound for \eqref{accelLeader} and \eqref{accelFollower} while the maximum braking force is applied}.

The leader vehicle cost function $J_L$ penalizes deviations from the desired velocity $v_L^{des}$ \eqref{leaderObj1}, nonzero control inputs \eqref{leaderObj2}, and nonzero control input rates \eqref{leaderObj3}, effectively penalizing vehicle jerk. The scalar $\alpha \in \R_{> 0}$ and matrix $R \in \R^{2 \times 2}$ are design parameters which allow one to tune controller performance. Increasing $\alpha$, for example, smooths the acceleration and deceleration profiles of the vehicle, but reduces the controller's agility. Furthermore, we set
\begin{equation}
R = \begin{bmatrix}
R_a & R_0 \\
R_0 & R_b
\end{bmatrix}
\end{equation}
where the diagonal entries $R_a$, $R_b \in \R$ can be increased to encourage the controller to use smaller actuation torques $T_{a,ref}^L(t)$ and $T_a^L(t)$, respectively, and the off-diagonal entries $R_0 \in \R$ are made sufficiently large in order to prevent the accelerating and braking control inputs from being active simultaneously.

The leader MPC problem is subject to the following constraints: vehicle dynamics \eqref{leaderConstr1}, lower and upper bounds on velocity \eqref{leaderConstr2}, distance constraint \eqref{leaderConstr3}, torque and reference torque constraints \eqref{leaderConstr4} - \eqref{leaderConstr6}, and initial condition \eqref{leaderConstr7}. The terminal constraint \eqref{leaderConstr8} ensures the leader maintains a safe distance to any obstacle ahead (namely, a front vehicle or intersection requiring a stop), and will be discussed further in Section \ref{safetyUrban}. The parameters $d_{\min}$ and $t_h$ are tuned to increase passenger comfort. For example, if $t_h$ is too small it may feel as if the vehicle is braking late when approaching slow-moving traffic or a stop bar, and if $t_h$ is too large the vehicle will brake harshly in response to cut-in vehicles. The values of all MPC parameters are given in Table \ref{MPCParameters}.

\addedTwo{At each time step, the leader vehicle solves its MPC problem and obtains an optimal control input sequence and velocity trajectory:
\begin{align}
& u_L(t | t), \ u_L(t + 1 | t), \ \dots, \ u_L(t + N_p | t), \label{inputSequence} \\
& v_L(t | t), \ v_L(t + 1 | t), \ \dots, \ v_L(t + N_p + 1 | t). \label{velSequence}
\end{align}
The first control input $u_L(t | t)$ of the sequence \eqref{inputSequence} is then implemented on the vehicle, and the MPC problem is solved again at the next time step. Furthermore, the computed velocity trajectory in \eqref{velSequence} is sent to the other platooning vehicles at each time step via V2V communication as a velocity forecast, as discussed in Section \ref{v2vcomm}}.

\subsection{Follower vehicle MPC} \label{followMPC}
The goal of each follower vehicle is to maintain a desired distance $s_i^{\text{des}}$ to the leader vehicle, while also maintaining a minimum safety distance $d_{min}$ to the front vehicle at all times. The MPC problem to be solved is defined as follows
\begin{subequations}
\begin{align}
& \underset{u_i(\cdot|t)}{\text{min}}
& J_i & = \sum_{k=t}^{t+N_p+1} (s_i(k|t) - s_i^{des})^2 \label{followerObj1} \\
& & & + \sum_{k=t}^{t+N_p} u_i(k|t)^T R u_i(k|t) \label{followerObj2} \\
& & & + \alpha \sum_{k=t}^{t+N_p-1} \|u_i(k+1|t) - u_i(k|t)\|^2 \label{followerObj3} \\
& \text{s.t.} & & x_i(k+1|t) = \label{followerConstr1} \\
& & & \quad A_i x_i(k|t) + B_i u_i(k|t) + E_i \hat{w}_i(k), \nonumber \\
& & & v_{\min} \leq v_i(k|t) \leq v_{\max}, \label{followerConstr2} \\
& & & d_{\min} \leq h_i(k|t), \label{minDistanceMPC} \\
& & & 0 \leq T_i^a(k|t) \leq T_{\max}^a, \label{followerConstr3} \\
& & & 0 \leq T_i^{a, ref}(k|t) \leq T_{\max}^a, \label{followerConstr4} \\
& & & 0 \leq T_i^b(k|t) \leq T_{\max}^b, \label{followerConstr5} \\
& & & x_i(t|t) = \hat{x}_i(t), \label{followerConstr6} \\
& & & \forall k = t, \dots, t+N_p, \nonumber \\
& & & \begin{bmatrix}
h_i(t+N_p|t) \\ v_i(t+N_p|t)
\end{bmatrix} \in C_F(\hat{v}_{i - 1}(t+N_p)), \label{followerConstr7}
\end{align}
\end{subequations}
where the notation used is the same as in \eqref{leadMPCproblem}. The follower vehicle objective function $J_i$ penalizes deviations from the desired distance to the leader vehicle, given by
\begin{equation}
s_i^{\text{des}} := d^{\text{des}} \cdot i,
\end{equation}
where $d^{\text{des}}$ is a design parameter. Furthermore, we also include penalties on input \eqref{followerObj2} and jerk \eqref{followerObj3}. Similar to the leader, these penalties have to be adjusted carefully to balance performance and passenger comfort. Furthermore, we note that constraints \eqref{followerConstr2} and \eqref{followerConstr7} are imposed with respect to the front (platooning) vehicle only, since safety tasks regarding an upcoming intersection are handled by the platoon leader (the terminal constraint \eqref{followerConstr7} will be discussed further in the next section).

\addedTwo{Similar to the leader, at each time step the follower vehicle solves its MPC problem and obtains an optimal control input sequence and velocity trajectory. We apply the first control input of the sequence, and the computed velocity trajectory is broadcast to the platoon via V2V communication. Hence, since velocity forecasts \eqref{velForecasts} are received by all follower vehicles via V2V communication, we use the following disturbance preview for MPC:
\begin{align}
\hat{w}_i(k) &:= [\hat{v}_L(k); \ \hat{v}_{i-1}(k)] \nonumber \\
& \ = [v_L(k|t); \ v_{i-1}(k|t)], \quad  k = t, \dots, t + N_p, \label{velEstimates}
\end{align}
where the planned velocity trajectories $v_L(k|t)$ and $v_{i-1}(k|t)$ were computed by the leader and front vehicle when they solved their respective MPC problems. 
}
\addedTwo{
\begin{remark} \label{trustRemark}
Since we use the \textit{full} velocity forecast as a disturbance preview in \eqref{velEstimates}, a natural question that arises is whether or not these predictions are reliable. To address this question, in \cite{smith2019balancing} we defined the \textit{trust horizon} $F$, which allows us to adjust how much of the velocity forecasts are used. For a trust horizon of $F$, time steps $t$ through $t + F$ of all velocity forecasts are used. After time step $t + F$ the front vehicle is assumed to decelerate at the maximum rate until coming to a stop, and therefore the terminal constraint \eqref{followerConstr7} is imposed at time step $t + F$. This is in contrast to the approach in this paper, where we assume the front (platooning) vehicle will fully realize the trajectory in its velocity forecast as in \eqref{velEstimates}, corresponding to $F = N_p$. Doing so introduces some risk to the follower vehicles; however, this is necessary to achieve a reasonable increase in traffic throughput with vehicle platooning, as shown in our previous work \cite{smith2019balancing}.
\end{remark}}

\begin{table}
    \caption{MPC Parameters}
    \label{MPCParameters}
    \centering
    \begin{tabular}{c l l l}
    \toprule
         $d^{\text{des}}$ & desired distance & m & 6 \\
         $d_{\min}$ & minimum distance (front vehicle) & m & 6 \\
         $d_{\min}$ & minimum distance (stop bar) & m & 5 \\
         $t_h$ & time headway & s & 1.6 \\
         $v_L^{\text{des}}$ & desired velocity (leader) & m$/$s & 15 \\
         $v_{\min}$ & minimum velocity & m$/$s & 0 \\
         $v_{\max}$ & maximum velocity & m$/$s & 20 \\
         $T_{\max}^a$ & maximum accelerating torque & Nm & 1500 \\
         $T_{\max}^b$ & maximum braking torque & Nm & 2000 \\
         $N_p$ & MPC horizon & - & 20 \\
    \bottomrule
    \end{tabular}
\end{table}

\section{Safety Constraints and MPC Solution} \label{safetyUrban}
We now discuss how we formally ensure safety in an urban traffic setting. First, in Section \ref{priorityTarget} we describe the set of safe states for a vehicle in relation to the two primary obstacles it can encounter in an urban setting: another vehicle ahead of it, and an upcoming intersection. Furthermore, we show that at each time instant the vehicle needs to consider only one of these obstacles, which we refer to as the \textit{priority obstacle}, thereby simplifying the task of ensuring safety. Next, in Section \ref{efficientSolution} we discuss how we use the safe sets from Section \ref{priorityTarget} in our MPC problems, as well as how we efficiently solve the MPC problems at runtime.

\begin{figure}
    \centering
    
    \begin{subfigure}[b]{\columnwidth}
        \input{tikz/terminal_set_lead.tex}
        \caption{$\mathcal{C}_{F}(v_F(0))$ for $v_F(0) = 14$m/s and $d_{\min} = 6$m.}
        \label{fig:terminal1}
    \end{subfigure}
    
    \vspace{0.2cm}
    
    \begin{subfigure}[b]{\columnwidth}
        \input{tikz/terminal_set_int}
        \caption{$\mathcal{C}_{TL}$ for $d_{\min} = 5$m.}
        \label{fig:terminal2}
    \end{subfigure}
    
    \caption{In \ref{fig:terminal1} and \ref{fig:terminal2} we plot the terminal sets \eqref{frontSet} and \eqref{intersectionSet} for the front vehicle and upcoming intersection, respectively. For computing the sets, we use $a_{\text{min,brake}} = 3.2$ m/s$^2$ and $a_{\text{max,brake}} = 5.0912$ m/s$^2$.}
    \label{fig:terminal}
\end{figure}
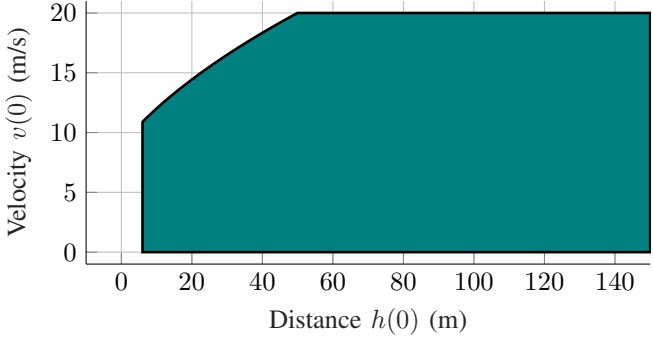
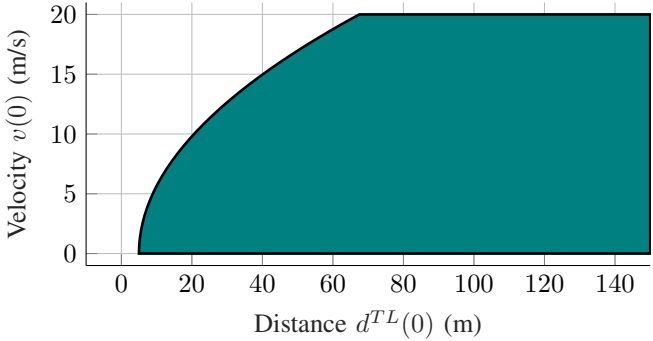

\subsection{Safe States and Priority Obstacle} \label{priorityTarget}
Consider an ego vehicle (representing either a platoon leader or follower here), a front vehicle ahead of it, and an upcoming intersection. Throughout the section, we let $a(t)$ and $a_F(t)$ be the accelerations of the ego and front vehicles, respectively, so that the vehicle dynamics become
\begin{align}
\dot{h}(t) &= v_F(t) - v(t), \nonumber \\
\dot{d}^{TL}(t) &= -v(t), \nonumber \\
\dot{v}_F(t) &= a_F(t), \nonumber \\
\dot{v}(t) &= a(t), \label{kinematic}
\end{align}
where $h(t)$ is the headway of the ego vehicle, $d^{TL}(t)$ is the distance from the ego vehicle to the upcoming traffic light stop bar, and $v_F(t)$ and $v(t)$ are the velocities of the front and ego vehicles, respectively. Since we observed no actuation delay while braking during experimentation, it is sufficient to use \eqref{kinematic} in place of \eqref{leadDyn} for the analysis here.

We first assume that only a front vehicle is present, and define safety for the ego vehicle with respect to the front vehicle as
\begin{equation}
h(t) \geq d_{\min}, \quad t \geq 0. \label{frontSafety}
\end{equation}
To enforce \eqref{frontSafety}, the ego vehicle must ensure it can maintain a minimum safety distance $d_{\min}$ if the front vehicle applies the maximum braking force until coming to a stop. We formalize this requirement in the following Proposition:
\begin{proposition} \label{prop1}
Consider the vehicle dynamics given in \eqref{kinematic}. Let $a_{\text{min,brake}}$, $a_{\text{max,brake}} \in \R_{>0}$, and $a_{\text{min,brake}} \leq a_{\text{max,brake}}$. Suppose the accelerations $a_F(t)$ and $a(t)$ satisfy
\begin{align}
a_F(t) &= \begin{cases}
-a_{\text{max,brake}}, & t \in [0, t_F^s], \label{frontBraking} \\
0, & t > t_F^s,
\end{cases} \\[5pt]
a(t) &= \begin{cases}
-a_{\text{min,brake}}, & t \in [0, t^s], \label{leadBraking} \\
0, & t > t^s,
\end{cases}
\end{align}
where $t_F^s := v_F(0) / a_{\text{max,brake}}$ and $t^s := v_L(0) / a_{\text{min,brake}}$ are the first time instants in seconds such that $v_F(t_F^s) = 0$ and $v(t^s) = 0$, respectively. Then, \eqref{frontSafety} will hold if $[h(0); v(0)] \in \mathcal{C}_F(v_F(0))$, where
\begin{align}
& \mathcal{C}_F(v_F(0)) := \label{frontSet} \\
& \quad \left\{
\begin{bmatrix} h(0) \\ v(0) \end{bmatrix} : 
\quad \begin{aligned}
& h(0) \geq \frac{v(0)^2}{2 a_{\text{min,brake}}} - \frac{v_F(0)^2}{2 a_{\text{max,brake}}} + d_{\min}, \\[5pt]
& h(0) \geq d_{\min}, \quad v(0) \geq 0
\end{aligned} \right\} \nonumber
\end{align}
\end{proposition}
for $v_F(0) \in \R_{\geq 0}$. For a proof we refer to \cite{turri2015fuel}, Lemma 1 (see also \cite{li1997ahs, shalev2017formal}). We note that in addition to $v_F(0)$, the set $\mathcal{C}_F(v_F(0))$ also depends on $a_{\text{min,brake}}$, $a_{\text{max,brake}}$, $d_{\min} \in \R_{>0}$. A plot of $\mathcal{C}_F$ is given in Figure \ref{fig:terminal1}.

Next, we suppose that only an upcoming intersection requiring a stop is present. In this case, the ego vehicle must ensure it can make a complete stop and leave a distance of $d_{\min}$ to the intersection stop bar. Formally, we require that if the ego vehicle decelerates until coming to a stop as in \eqref{leadBraking}, then the following will hold
\begin{equation}
d^{TL}(t) \geq d_{\min}, \quad t \geq 0. \label{intersectionSafety}
\end{equation}
We note that when the light cycles to green, this constraints is relaxed and the platoon is allowed to proceed. Similar to Proposition \ref{prop1}, we can show that \eqref{intersectionSafety} holds if the ego vehicle decelerates as in \eqref{leadBraking} and $[d^{TL}(0); \ v(0)] \in \mathcal{C}_{TL}$, where
\begin{align} \label{intersectionSet}
\mathcal{C}_{TL} := \left\{
\begin{bmatrix} d^{TL}(0) \\ v(0) \end{bmatrix} :
\; \begin{aligned}
d^{TL}(0) &\geq \frac{v(0)^2}{2 a_{\text{min,brake}}} + d_{\min}, \\[5pt]
v(0) &\geq 0
\end{aligned} \right\}.
\end{align}
A plot of $\mathcal{C}_{TL}$ is given in Figure \ref{fig:terminal2}.

\begin{figure}
    \centering
    
    \begin{subfigure}[b]{\columnwidth}
        \includegraphics[width = \linewidth]{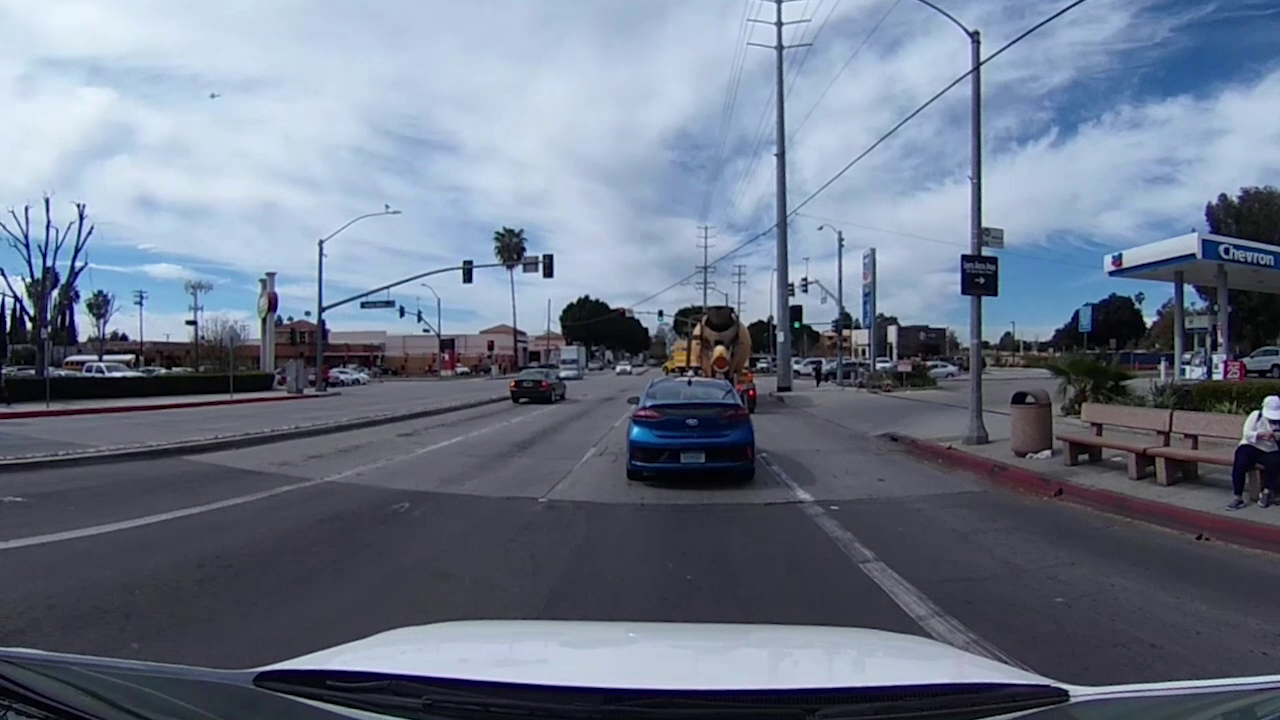}
        \caption{Truck has priority.}
        \label{fig:priority1}
    \end{subfigure}
        
    \vspace{0.2cm}
    
    \begin{subfigure}[b]{\columnwidth}
        \includegraphics[width = \linewidth]{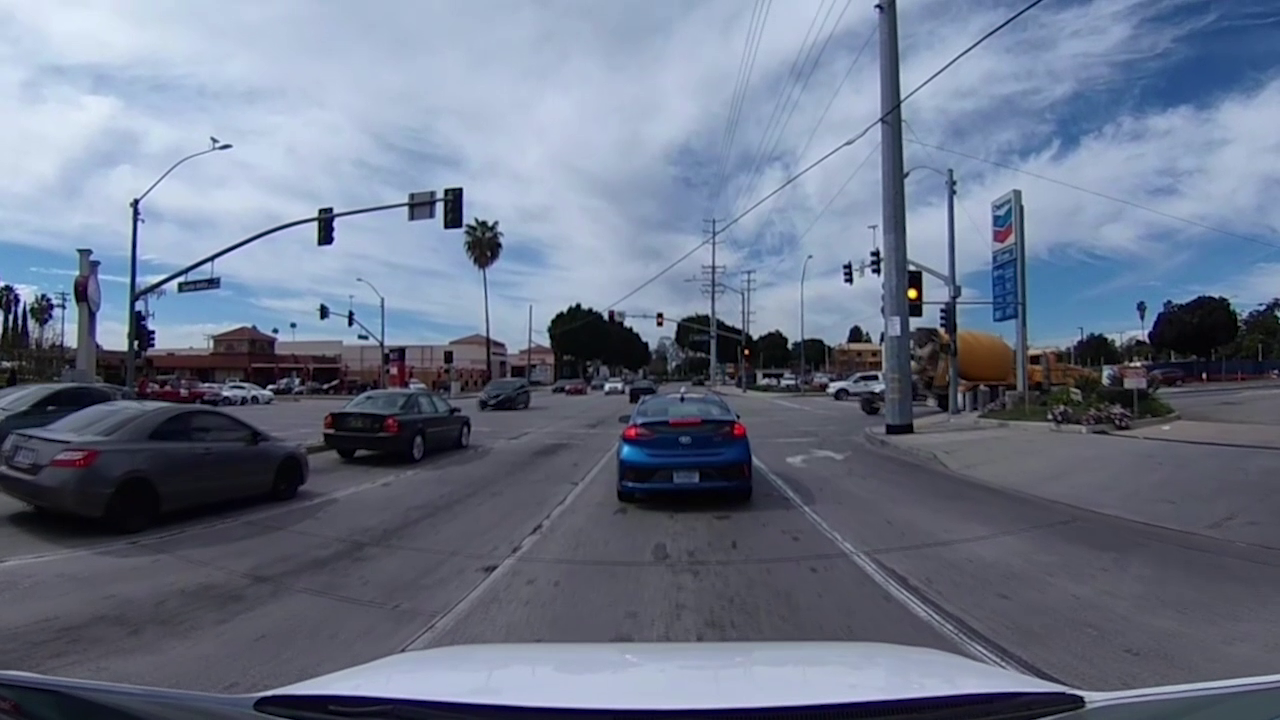}
        \caption{Intersection has priority.}
        \label{fig:priority2}
    \end{subfigure}
    
    \caption{View from the middle platooning vehicle as it approaches an intersection during our demonstration in Arcadia, CA. In Figure \ref{fig:priority1} there is a slow-moving truck attempting to turn right ahead of the leader vehicle. Since the truck takes priority over the intersection at this point, the platoon is forced to slow down. In Figure \ref{fig:priority2} the truck completes the right turn and priority shifts to the intersection.}
    \label{fig:priority}
\end{figure}

Now, we suppose that both a front vehicle and an upcoming intersection requiring a stop are present simultaneously. In this scenario, we require that if the front and ego vehicle (representing the platoon leader here) decelerate until coming to a stop as in \eqref{frontBraking} and \eqref{leadBraking}, then \textit{both} \eqref{frontSafety} and \eqref{intersectionSafety} will hold. To determine which obstacle is prioritized, the ego vehicle can check if the following condition holds:
\begin{equation} \label{targetCondition} 
h(0) + \frac{v_F(0)^2}{2 a_\text{max,brake}} \leq d^{TL}(0).
\end{equation}
If \eqref{targetCondition} holds then the front vehicle is capable of stopping in front of the intersection stop line, and therefore must be prioritized. If \eqref{targetCondition} does not hold then the upcoming intersection is prioritized (see Figure \ref{fig:priority} for an illustration). We summarize this idea in the following Proposition, which follows directly from the definitions of $\mathcal{C}_F$ and $\mathcal{C}_{TL}$.
\begin{proposition} \label{prop2}
Assume $h(0) \geq d_{\min}$. If \eqref{targetCondition} does not hold, then $[d^{TL}(0); \ v(0)] \in \mathcal{C}_{TL}$ implies $[h(0); \ v(0)] \in \mathcal{C}_F(v_F(0))$. Otherwise, if \eqref{targetCondition} holds, then $[h(0); \ v(0)] \in \mathcal{C}_F(v_F(0))$ implies $[d^{TL}(0); \ v(0)] \in \mathcal{C}_{TL}$. 
\end{proposition}
Based on Proposition \ref{prop2}, we conclude that for the leader vehicle MPC problem discussed in the previous section, it is sufficient to impose a terminal constraint with respect to only the priority obstacle. This is beneficial for efficiently solving the MPC problems at runtime, as discussed further in the next section.

\begin{remark}
In the above discussion we assumed $d_{\min}$ is the same for both the front vehicle and the intersection, whereas in our experiments we used slightly different values of $d_{\min}$ for each. Although this is beneficial for passenger comfort, there is one drawback to this adjustment: in corner cases where priority between the two obstacles can easily switch, we may only satisfy \eqref{frontSafety} and \eqref{intersectionSafety} for the minimum of these two values, i.e. for $d_{\min} := \min\{ d_{\min,F}, \ d_{\min,TL} \}$, where $d_{\min,F}$ and $d_{\min,TL}$ are the unique minimum distance values used for the front vehicle and intersection, respectively. We ensured, however, that this minimum safety margin is still sufficient for testing purposes. Furthermore, in normal traffic conditions the priority between obstacles is clear (usually, the front vehicle is clearly stopping at the intersection, or clearly passing through it).
\end{remark}

\begin{remark}
If an upcoming intersection is not present (or does not require a stop), then the front vehicle is prioritized if one is present. This allows, for example, the platoon to pass through a green light if it is safe to do so. Similarly, if only a front vehicle is present then it is prioritized. If neither obstacle is present, then no obstacle-related constraints are imposed on the leader.
\end{remark}

\subsection{Terminal Constraints and MPC Solution} \label{efficientSolution}

We now connect the discussion in the previous section to terminal constraints \eqref{leaderConstr8} and \eqref{followerConstr7}. For the follower vehicles, the primary safety task is to maintain a minimum distance to the front (platooning) vehicle. Therefore, the terminal constraint \eqref{followerConstr7} is imposed with respect to the front vehicle only. For the leader vehicle, the primary safety tasks are to stop at an upcoming intersection when necessary, and to maintain a minimum distance to the front (non-platooning) vehicle. Based on the discussion in Section \ref{priorityTarget}, this is accomplished by imposing the terminal constraint \eqref{leaderConstr8} with respect to the priority obstacle. To this end, we define
\begin{align}
& d_L^*(t + k | t) := \label{dStar} \\
& \quad \begin{cases}
h_L(t + k | t), & \text{if } \hat{x}_L(t) \text{ and } \hat{v}_{F}(t) \text{ satisfy } \eqref{targetCondition}, \\
d_L^{TL}(t + k | t), & \text{otherwise,}
\end{cases} \nonumber
\end{align}
as the planned distance from the leader to the priority obstacle at time step $k$, computed at time step $t$, and
\begin{align}
& \mathcal{C}(\hat{x}_L(t), \hat{v}_F(t), \hat{v}_F(t+N_p)) := \nonumber \\
& \quad \begin{cases}
\mathcal{C}_F(\hat{v}_F(t + N_p)), & \hat{x}_L(t) \text{ and } \hat{v}_F(t) \text{ satisfy } \eqref{targetCondition}, \\
\mathcal{C}_{TL}, & \text{otherwise},
\end{cases}
\end{align}
as the terminal set with respect to the priority obstacle. We note that the priority obstacle will be the same throughout the MPC planning horizon, since \eqref{targetCondition} checks whether the front vehicle can stop before the intersection stop bar if it decelerates at the rate $a_{\text{max,brake}}$, which is its assumed behavior in the leader MPC problem in \eqref{frontBrakingMPC}.

\begin{figure*}
\centering

\input{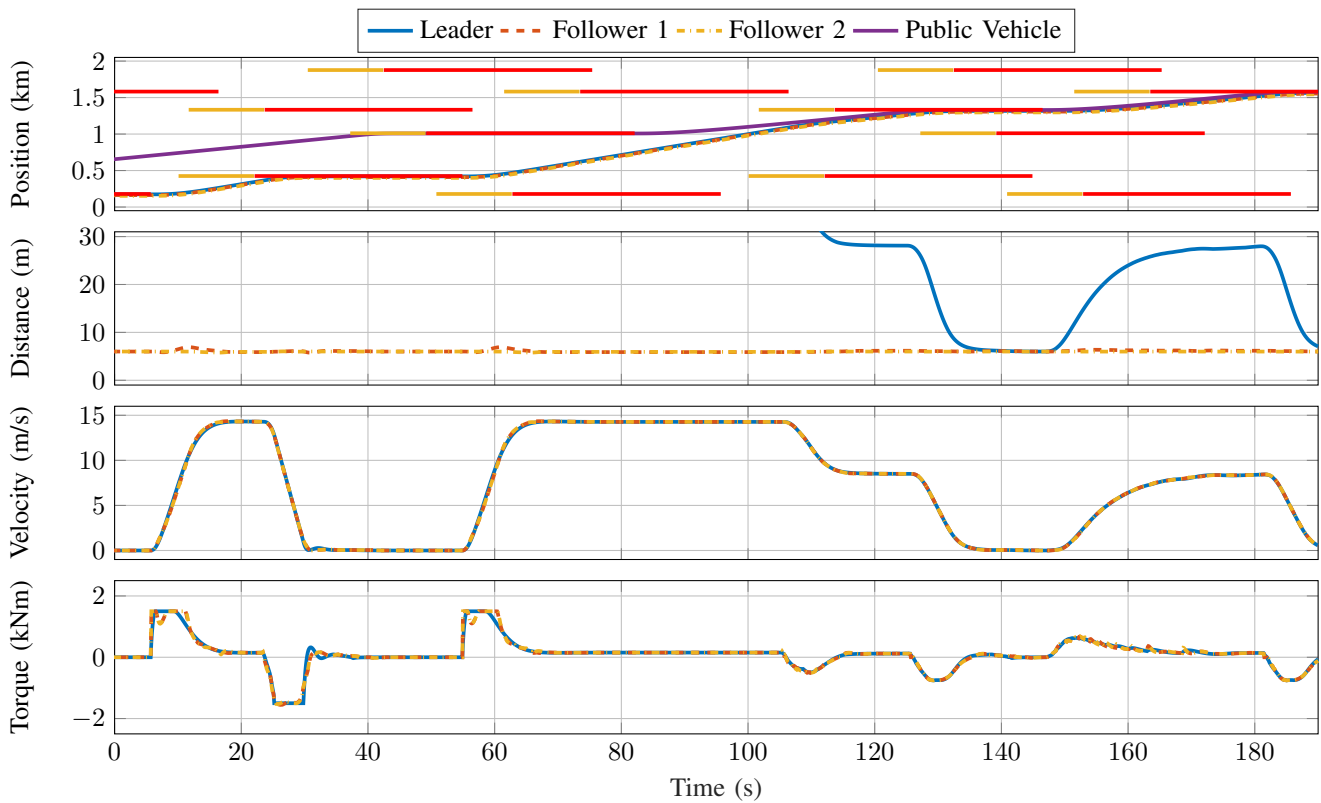}
\caption{Simulation results for an urban traffic scenario with a non-platooning lead vehicle and multiple signalized intersections. In the top plot, we show the position of all simulated vehicles (including the public vehicle which is not platooning), as well as the position of each intersection which has either a yellow or red phase. In the bottom three plots, we show the inter-vehicle distances (including the distance from the leader to the public vehicle), velocities, and torque commands for the platooning vehicles. \label{fig:arcadiaSim}}

\end{figure*}

To solve the leader and follower vehicle MPC problems at runtime we use the tool CVXGEN \cite{mattingley2012cvxgen}, which allows one to generate C code for solving a custom quadratic program (QP) reliably and efficiently. Since CVXGEN can only be used for moderately-sized QPs, it is beneficial to impose terminal constraint \eqref{leaderConstr8} with respect to only the priority obstacle, as imposing a terminal constraint with respect to both obstacles would create additional (redundant) constraints. Furthermore, since our MPC problems must be represented as QPs with linear constraints, the sets $\mathcal{C}_F$ and $\mathcal{C}_{TL}$ discussed in the previous section cannot be directly encoded into our MPC problems. Instead, we use a procedure from \cite{lefevre2015learning} to compute polyhedral constraint sets to be used in place of $\mathcal{C}_F$ and $\mathcal{C}_{TL}$. In particular, we compute a collection of sets $\mathcal{C}_F(v_F(0))$ to be used for $v_F(0) \in [v_{\min}, v_{\max}]$. This collection of sets is computed offline, and the proper set is selected during runtime to be used for MPC (for more details, we refer the reader to \cite{smith2019balancing}).

Since it is important to avoid infeasibility of the MPC problems during experimentation, all constraints in each problem (except for the vehicle dynamics constraints) are converted to soft constraints. This means that for a hard constraint such as $G x \leq h$, where $x \in \R^n$, $G \in \R^{m \times n}$, and $h \in \R^m$, we instead add the term $\lambda \mathbb{1}^T (G x - h)_+$ to the objective function, where $\lambda \in \R_{> 0}$, $\mathbb{1} \in \R^{m}$ is the vector of all 1's, and $y_+$ for $y \in \R^{m}$ indicates that we are thresholding each element of $y$ so that $y_+ \in \R^{m}_{\geq 0}$ (see \cite{CVXGENinfeasibility}).

\section{Simulation Results} \label{simulation}
\added{We now present results from our simulation tool developed in MATLAB, which enabled us to validate the platooning software prior to conducting real-world experiments. In particular, the tool is useful to confirm that the platoon preserves safety even when it encounters traffic lights and other non-platooning vehicles, using the approach in Sections \ref{v2icomm} and \ref{safetyUrban}. Furthermore, we are able to estimate the potential gains in traffic throughput at intersections, using a metric from \cite{smith2019balancing}.}

\subsection{Urban Stop and Go Scenario} \label{stopGo}
We use our tool to simulate \addedThree{the vehicle platoon travelling along an arterial roadway with moderate traffic. In particular, our goal here is to imitate the conditions we will encounter during our field experiments in Arcadia, CA (see Section \ref{arcadia}). To simulate public vehicles in traffic, we create velocity trajectories in simulation and then replay them so that simulations are repeatable. Taking into account the positions / velocities of the platoon leader and a public vehicle ahead of it, we can send radar signals as an input to the leader controller and observe how the platoon responds. Furthermore, we can also create signalized intersections with the following attributes: position (m), V2I communication range (m), cycle offset (s), red / yellow / green time (s), and cycle length (s). We placed intersections along the simulated arterial road so that the distances between traffic lights are similar to the Arcadia corridor discussed in Section \ref{arcadia}. All the individual intersections are composed to create a traffic network object which can be queried to determine the nearest upcoming traffic light relative to the platoon leader. As the platoon leader approaches the intersection, we send V2I messages from that traffic light as an input to the leader controller and simulate the platoon response}.

The simulation results are shown in Figure \ref{fig:arcadiaSim}. \addedThree{In particular, we note that the horizontal yellow and red lines in the top plot represent intersections which have a yellow and red phase at that time instant, respectively. Furthermore, the purple line represents the position of a public vehicle which is not platooning}. In the beginning of the simulation, the platoon encounters red lights at the first few intersections, stopping at each. Near the end of the simulation the platoon approaches a \addedTwo{(non-platooning)} public vehicle which is travelling much more slowly, and the platoon is forced to reduce its speed for the remainder of the simulation. We note that near the end of the simulation, the public vehicle comes to a complete stop at an intersection and as a result the platoon leader also stops, leaving a distance of 6m as desired. \addedThree{As mentioned previously, one of the primary goals of the simulation tool is to verify that the platoon responds appropriately when it encounters other non-platooning vehicles and signalized intersections. Observing the simulation results, we can see that the platoon stops at each intersection when necessary, and that the distance from the leader to the public vehicle stays above 6m at all times as desired}. Lastly, we remark that for simulation we did not use the same controller parameters that we did for experimentation, where the parameters were mainly selected to improve passenger comfort.

\begin{table}

    \centering

    \caption{Improved Throughput \label{tab:throughputAnalysis1}}
    \begin{tabular}{l l}
    \toprule
         Simulation Intersection 1 & 4,336.4 vph \\
         Simulation Intersection 2 & 4,336.4 vph \\
         Simulation Intersection 4 & 2,477.8 vph \\
         Test Track Intersection (Virtual) & 4,463.4 vph \\
    \bottomrule
    \end{tabular}
    
    \bigskip
    
    \caption{Baseline Throughput \label{tab:throughputAnalysis2}}
    \begin{tabular}{l l}
    \toprule
         Simulation Intersection 1 & 2149.8 vph \\
         Simulation Intersection 2 & 2156.9 vph \\
         Simulation Intersection 4 & 1710.5 vph \\
         Test Track Intersection (Virtual) & 2730.7 vph \\
    \bottomrule
    \end{tabular}
    
\end{table}

\subsection{Estimating Throughput} \label{throughput}
\added{We now analyze the performance of the vehicle platooning system by estimating intersection throughput. To do so, we recall a performance metric defined in \cite{smith2019balancing}. At time $t = 0$ let the platoon be stopped at the (current) intersection stop bar with no vehicles ahead
\begin{align*}
[p_L(0); \ v_L(0)] &= [-d_{\min}; \ 0], \\
[p_i(0); \ v_i(0)] &= [-d_{\min} - (L_{\text{veh}} + d^{\text{des}}) \cdot i; \ 0], \\
& \qquad \quad i = 1, \dots, N - 1,
\end{align*}
where $L_{\text{veh}}$ is the vehicle length (assumed to be uniformly 4.5 meters for all vehicles), and the intersection stop bar is assumed to be positioned at 0 meters. Suppose at time $t = 0$ the traffic light cycles from red to green, and the platoon immediately starts moving through the intersection. Let $\ell \in \R_{> 0}$ be the length of the intersection in meters, and define $t_L$ and $t_{N-1}$ to be the smallest time instants in seconds such that $p_L(t_L) \geq \ell$ and $p_{N-1}(t_{N-1}) \geq \ell$, respectively. We then estimate intersection throughput in vehicles per hour as
\begin{equation} \label{throughputEstimate}
\text{throughput (vph)} \approx 3600 \cdot \frac{N - 1}{t_{N-1} - t_L}.
\end{equation}
Thus, performance is maximized when the platoon 1) accelerates to a high velocity while crossing the intersection, and 2) accurately maintains the desired inter-vehicle gaps while accelerating. We note that for the estimate \eqref{throughputEstimate} to be accurate, we must consider the length of each vehicle, as opposed to treating each as a point mass.}

\added{Throughput analysis of simulation results (as well as the test-track experiments discussed in Section \ref{CPG}) is shown in Tables \ref{tab:throughputAnalysis1} and \ref{tab:throughputAnalysis2}, where all estimates are obtained via \eqref{throughputEstimate}. In particular, throughput is estimated at the 1st, 2nd, and 4th intersection, located at approximately 0.18 km, 0.43 km, and 1.33 km in the simulation, respectively. \addedTwo{In Table \ref{tab:throughputAnalysis1} we show improved levels of throughput achieved using our vehicle platooning system, which are estimated from the simulation run shown in Figure \ref{fig:arcadiaSim}. In Table \ref{tab:throughputAnalysis2} we show baseline levels of throughput, which are estimated by running the same simulation with the \textit{trust horizon} (discussed in Remark \ref{trustRemark}) set to $F = 0$}. We note that throughput is much lower at the 4th intersection, due to the presence of a slower-moving public vehicle ahead of the platoon. Indeed, in situations like this, the benefit of vehicle platooning in terms of traffic throughput may not be fully realized.} \addedThree{We note also that our predictions here are in line with predictions from previous works which utilized simulation. For example, in \cite{lioris2017platoons} the authors predict that vehicle platooning could enable a saturation flow rate of 4800 vph per intersection movement}.

\begin{figure}
    \centering
    \includegraphics[width = \columnwidth]{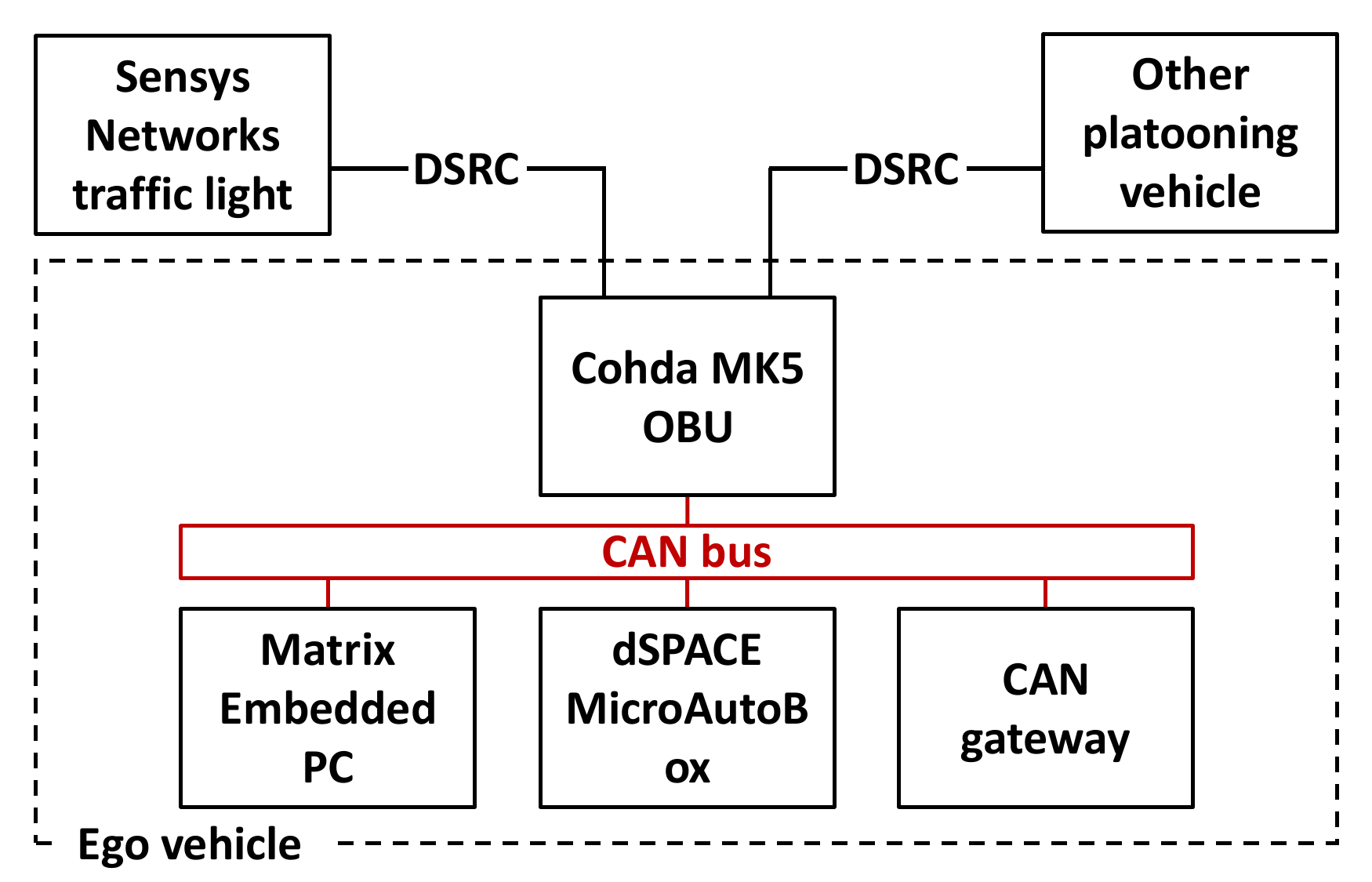}
    \vspace{1mm}
    \caption{Depiction of the on-board hardware setup for the test vehicles. The local CAN bus (in red) connects the computational devices (Matrix embedded PC and dSPACE MicroAutoBox) to the Cohda OBU for DSRC communication. The HCU (CAN gateway) provides an interface between the local CAN bus and the production systems of the test vehicle. Using the local CAN bus and the gateway functionality of the HCU, we can send commands and access measurements to and from the production systems without needing access to proprietary vehicle data.}
    \label{fig:hardware}
\end{figure}

\begin{figure*}
\input{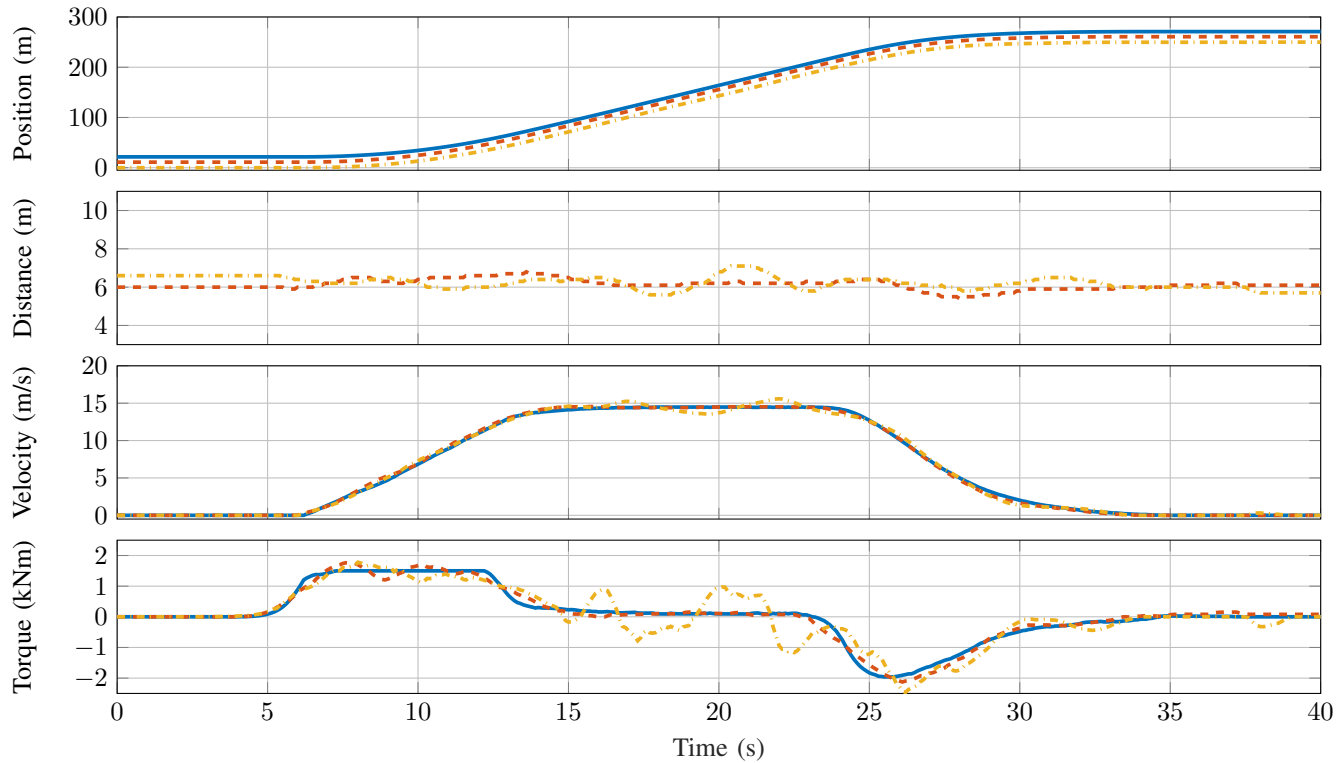}
\caption{Experimental results from the Hyundai-KIA Motors California Proving Grounds with the test vehicles shown in Figure \ref{fig:ioniqs}. Here, we had the platoon track a reference trajectory which was generated via our simulation tool. The position, inter-vehicle distance, velocity, and MPC torque command for each vehicle are shown in each subplot, respectively. The desired distance between vehicles was 6 meters. \label{fig:CPGtest}}
\end{figure*}

\section{Experimental Results} \label{experimentalResults}
\added{In this section we present the experimental results and evaluate the performance of our platooning controller via the throughput metric from Section \ref{throughput}. \addedTwo{We discuss the experimental setup in Section \ref{hardware}}, and in Section \ref{CPG} we present results from preliminary tests on a closed track at the Hyundai-KIA Motors California Proving Grounds in California City, CA. Next, we give an overview of a final platooning demonstration on public roadways in Arcadia, CA in Section \ref{arcadia}. Links to drone videos of each series of tests are also provided}.

\subsection{Test Vehicles} \label{hardware}
We use the three test vehicles shown in Figure \ref{fig:ioniqs}, each of which is equipped with a production forward-looking radar and camera that estimate the front vehicle distance, velocity, and acceleration. To enable V2V and V2I communication, we use a Cohda Wireless MK5 V2X on-board unit (OBU), which also has an integrated GPS. The Cohda OBU allows the vehicles to exchange BSMs and custom V2V messages, which include a velocity forecast and other information. This transmitted information allows the third vehicle in the platoon, for instance, to estimate its current distance to the leader vehicle. The Cohda also allows each vehicle to communicate with any nearby traffic lights which are instrumented to broadcast SPaT and MAP messages. Lastly, the controller for each vehicle is implemented on a dSpace MicroAutoBox, and a Matrix embedded PC exchanges information between the Cohda, MicroAutoBox, and the ego vehicle controller area network (CAN bus). The Matrix also runs a state machine which manages the role of each vehicle in the platoon, and is discussed further in Section \ref{statemanager}. A diagram of the hardware setup is shown in Figure \ref{fig:hardware}.

\added{An important hardware consideration for platooning is that of communication latencies}. In \cite{smith2019balancing} we discussed how including a time stamp in transmitted messages enables each vehicle to account for V2V communication delays. The idea is to use the time stamp to estimate the delay $d$ in time-steps (with sampling time $\Delta t = 0.1$s), and then to shift the velocity forecast used for MPC by $d$ steps, where we assume the transmitting vehicle will maintain a constant velocity beyond its planned trajectory. For the experimental work presented in this paper, however, we assume there are no communication delays between vehicles, which is done for two reasons. The first reason is that we have observed that communication latencies are typically small enough to be ignored for our application. The second reason is that estimating $d$ accurately is challenging in practice. Since the clocks on the test vehicle computers are not synchronized, one must estimate the clock skew between vehicles, which could potentially be time-varying, in order to accurately estimate delays.

\subsection{Closed track experiments} \label{CPG}
Preliminary vehicle platooning experiments were conducted on a closed test track at the Hyundai-KIA Motors California Proving Grounds in California City, CA (see Figure \ref{fig:ioniqs}). For all of the tests the leader vehicle does velocity tracking of a predetermined velocity trajectory (meaning $v_L^{des}$ in \eqref{leaderObj1} becomes time-dependent), and the follower vehicles do distance tracking relative to the leader vehicle. The predetermined velocity trajectories used for tracking were either from real velocity data collected during previous experiments, or artificial velocity data generated by our simulation tool. In Figure \ref{fig:CPGtest} we show experimental results from a test using artificial velocity data which has a step function-like trajectory. For these experiments we used a larger admissible range of the wheel torque for the follower vehicles, as seen in the bottom plot of Figure \ref{fig:CPGtest}. \addedThree{In particular, we note that the torque plotted is the \textit{desired} torque, i.e. the output of the MPC algorithm, as opposed to the \textit{measured} torque (estimated by the vehicle). However, the inter-vehicle distances and vehicle velocities are both from on-board measurements (the position data is then obtained offline by integrating the velocity data). We can see that as the platoon accelerates and decelerates, the followers accurately track the desired distance of 6m to the front vehicle - all tracking errors stay below about 1m throughout the experiment}. We note, however, that there is slightly larger tracking error (as well as larger variation of the wheel torque command) for the second follower in this experiment. We can mainly attribute this to state estimation error since GPS was used to estimate the distance $s_i(t)$ for all experiments at the California Proving Grounds, as discussed in Section \ref{stateEstimation}.

A video of the testing is available online at \url{https://youtu.be/U-O9iUZElR8}, which includes several test runs with varying levels of the \textit{trust horizon $F$} (discussed in Remark \ref{trustRemark}). We note that in test runs with a small trust horizon, for example $F = 10$ (half of the velocity forecast is trusted) or $F = 0$ (none of the velocity forecast is trusted, meaning the vehicles effectively do not use V2V communication), large gaps appear between the platooning vehicles while they are accelerating. This behavior is expected, since using the full velocity forecast relaxes the constraints on following distance so that the follower vehicles can get closer to the vehicle ahead. In the test run shown in Figure \ref{fig:CPGtest} we used $F = 15$, demonstrating that we are able to get accurate tracking performance when using a large portion of the velocity forecast (elsewhere in the paper we use $F = N_p = 20$). \added{Similar to Section \ref{throughput}, we estimate throughput for the test run shown in Figure \ref{fig:CPGtest} by treating the platoon as if it begins stopped at an intersection - our estimate is shown in Table \ref{tab:throughputAnalysis1}}. \addedTwo{Furthermore, in Table \ref{tab:throughputAnalysis2} we show a baseline level of throughput computed using data from a test run with $F = 0$. As expected, significantly higher throughput is achieved by utilizing the velocity forecast}.

\subsection{Public Road Demonstration} \label{arcadia}
\begin{figure}
    \centering
    \includegraphics[width = \columnwidth]{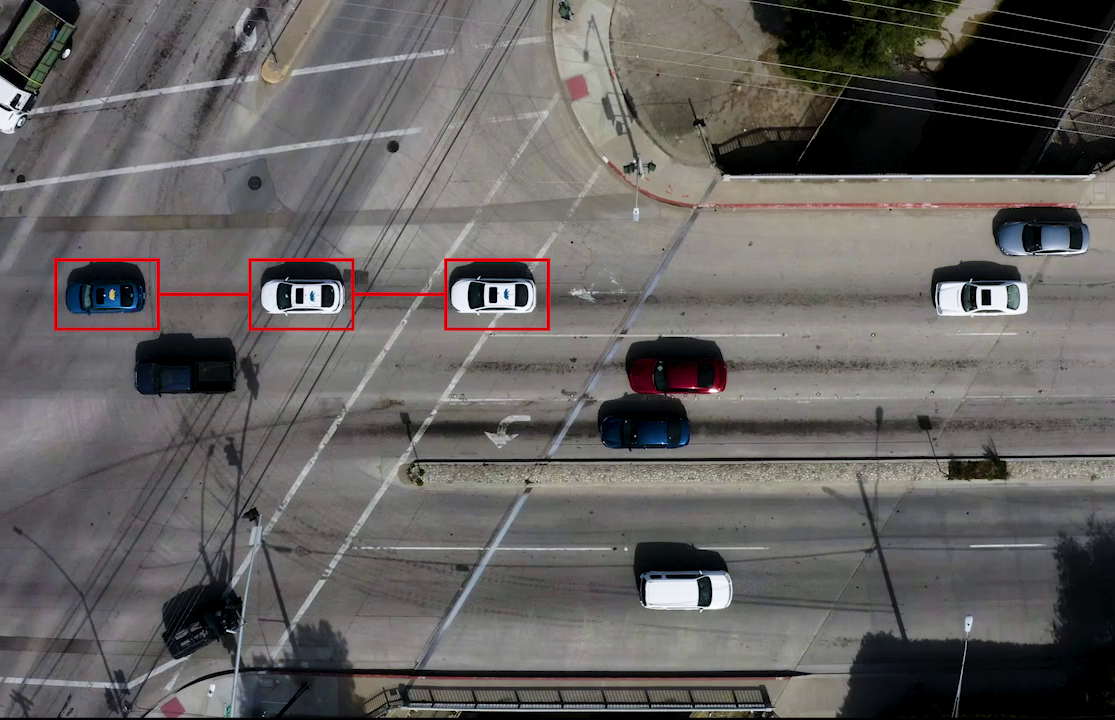}
    \caption{Overhead view of the platoon crossing an intersection in Arcadia, CA.}
    \label{fig:overhead}
\end{figure}

To demonstrate vehicle platooning in an urban environment with a moderate level of traffic, we conducted further experiments in Arcadia, CA. Our testing area is a 2.45 km long stretch of roadway on Live Oak Ave between S Santa Anita Ave and Peck Rd, and has eight consecutive intersections which are instrumented to send out SPaT and MAP messages for our vehicle platoon to receive. All tests in Arcadia were completed with a 3-vehicle platoon using the same MPC parameters as shown in Table \ref{modelParameters}, \added{with the exception that $v_L^{\text{des}} = 14$ m/s was used here}. Footage of our testing is available online: \url{https://youtu.be/xPYR_xP3FuY}. It captures a few instances where the platoon stops at the stop bar for a red light with no vehicles queued ahead of it. When the light turns green the platoon reacts immediately and moves through the intersection more quickly and compactly than the human-driven vehicles near it, further demonstrating the potential for throughput improvement (see Figure \ref{fig:overhead}).

\section{Conclusion and Future Work} \label{conclusion}
In this paper we presented the design and evaluation of a vehicle platooning system that can operate in an urban corridor with intersections and other traffic participants. The primary motivation for advancing platooning to an urban setting is to improve traffic efficiency by increasing throughput at intersections, which create bottlenecks for traffic flow. \added{We evaluated the performance of our vehicle platooning architecture by estimating the level of throughput that would be achieved at the intersection, calculated by measuring the time instants at which each vehicle crosses the intersection}.

An important challenge we encountered while testing on public roads is that of safely disengaging the platooning system and passing control back to the safety driver when necessary. To do so, we designed our system so that if any driver taps the brake pedal when the platoon is active, the controller for every platooning vehicle disengages immediately (via the finite state machine) and all drivers are notified immediately via a sound. We note, however, that this design can be problematic in certain scenarios. For example, suppose the platoon is approaching an intersection and begins braking when the driver in the leader vehicle, out of caution, disengages the platoon. This requires the drivers in the follower vehicles to react immediately, as their vehicles will suddenly stop braking when the controllers disengage. In the future we hope to address this issue by creating a safety system that ensures the vehicles start transitioning to a safe state immediately when the `plan' is cancelled, providing the safety driver more time to react. One potential approach, for example, is to have the platooning system transition to an ACC state of operation immediately after disengagement. The ACC system would then remain active and maintain a safe distance to the front vehicle until the driver takes over.

\added{Another future research direction relates to the procedure for setting cost weights in the MPC objective functions, which were manually tuned here. Indeed, in order to converge on acceptable values for the tuning parameters affecting vehicle drivability, such as the time headway constraint or penalty on vehicle jerk, multiple trial runs on a closed test track were necessary. To reduce development time, it would be interesting to see how a learning-based approach could potentially expedite this procedure. Furthermore, we note that the final tuning values we obtained are only valid for the class of test vehicles in this paper - other classes of vehicles, such as semi-trucks, have different performance characteristics and would therefore need separate tuning values. Learning-based performance tuning would also be beneficial for deploying a platoon with a large number of vehicles, since separate tuning values were used for each vehicle within the platoon in this paper. Learning could also accelerate the deployment of autonomous vehicles more broadly, by enabling companies to more easily meet customer's driving preferences}.

\ifCLASSOPTIONcaptionsoff
  \newpage
\fi

\section*{Acknowledgements}
We appreciate the funding received from the Department of Defense through the NDSEG Fellowship program and from the National Science Foundation through grant CNS-1545116, co-sponsored by the Department of Transportation. The authors would also like to thank Mikhail Burov, Galaxy Yin, and Emmanuel Sin for assisting with preliminary tests at the Richmond Field Station. Thanks also to Pravin Varaiya and Ching-Yao Chan for providing helpful suggestions throughout the project. The support of Hyundai America Technical Center and Sensys Networks is gratefully acknowledged.

% trigger a \newpage just before the given reference
% number - used to balance the columns on the last page
% adjust value as needed - may need to be readjusted if
% the document is modified later
%\IEEEtriggeratref{8}
% The "triggered" command can be changed if desired:
%\IEEEtriggercmd{\enlargethispage{-5in}}

% references section

% can use a bibliography generated by BibTeX as a .bbl file
% BibTeX documentation can be easily obtained at:
% http://mirror.ctan.org/biblio/bibtex/contrib/doc/
% The IEEEtran BibTeX style support page is at:
% http://www.michaelshell.org/tex/ieeetran/bibtex/
%\bibliographystyle{IEEEtran}
% argument is your BibTeX string definitions and bibliography database(s)
%\bibliography{IEEEabrv,../bib/paper}
%
% <OR> manually copy in the resultant .bbl file
% set second argument of \begin to the number of references
% (used to reserve space for the reference number labels box)
\bibliographystyle{IEEEtran}
\bibliography{IEEEabrv,bibfile}
\end{document}

%% file: tikz/terminal_set_lead.tex
% This file was created by matlab2tikz.
%
%The latest updates can be retrieved from
%  http://www.mathworks.com/matlabcentral/fileexchange/22022-matlab2tikz-matlab2tikz
%where you can also make suggestions and rate matlab2tikz.
%
\begin{tikzpicture}

\begin{axis}[%
width=7.5cm,
height=3.5cm,
at={(0cm,0cm)},
scale only axis,
xmin=-10,
xmax=150,
xlabel style={font=\color{white!15!black}},
xlabel={Distance $h(0)$ (m)},
ymin=-1,
ymax=21,
ylabel style={font=\color{white!15!black}},
ylabel={Velocity $v(0)$ (m/s)},
axis background/.style={fill=white},
axis x line*=bottom,
axis y line*=left,
xmajorgrids,
ymajorgrids
]

\addplot[area legend, line width=1.0pt, draw=black, fill=teal, forget plot]
table[row sep=crcr] {%
x	y\\
6               	0\\
6	                10.8977996167232\\
7.04259132284278	11.2000000001569\\
9.34659132283559	11.8400000001514\\
11.7785913223379	12.4800000000083\\
14.3385913224593	13.1200000000457\\
17.0265913198054	13.7599999994233\\
19.8425913249048	14.4000000005801\\
22.7865913228597	15.0400000001167\\
25.8585913225907	15.6800000000646\\
29.0585913218201	16.319999999917\\
32.3865913208938	16.9599999997413\\
35.8425913246574	17.6000000004332\\
39.4265913209092	18.2399999997533\\
43.1385913209124	18.8799999997679\\
46.9785913162414	19.5199999990086\\
49.946591322303	    20\\
150             	20\\
150             	0\\
% 48.9465913244603	19.8400000003451\\
% 45.0425913253275	19.2000000005106\\
% 41.2665913256606	18.5600000005794\\
% 37.6185913196096	17.9199999995233\\
% 34.0985913213553	17.2799999998274\\
% 30.7065913221595	16.6399999999826\\
% 27.4425913239538	16.0000000003398\\
% 24.3065913235914	15.3600000002709\\
% 21.298591319839	14.7199999994667\\
% 18.4185913219553	14.0799999999175\\
% 15.6665913227152	13.4400000001079\\
% 13.0425913234139	12.8000000002814\\
% 10.5465913209882	12.1599999996578\\
% 8.17859132226658	11.5199999999947\\
}--cycle;
\end{axis}
\end{tikzpicture}%

%% file: tikz/terminal_set_int.tex
% This file was created by matlab2tikz.
%
%The latest updates can be retrieved from
%  http://www.mathworks.com/matlabcentral/fileexchange/22022-matlab2tikz-matlab2tikz
%where you can also make suggestions and rate matlab2tikz.
%
\begin{tikzpicture}

\begin{axis}[%
width=7.5cm,
height=3.5cm,
at={(0cm,0cm)},
scale only axis,
xmin=-10,
xmax=150,
xlabel style={font=\color{white!15!black}},
xlabel={Distance $d^{TL}(0)$ (m)},
ymin=-1,
ymax=21,
ylabel style={font=\color{white!15!black}},
ylabel={Velocity $v(0)$ (m/s)},
axis background/.style={fill=white},
axis x line*=bottom,
axis y line*=left,
xmajorgrids,
ymajorgrids
]

\addplot[area legend, line width=1.0pt, draw=black, fill=teal, forget plot]
table[row sep=crcr] {%
x	y\\
5.00000000001091	0\\
5.01600000002145	0.320000000218155\\
5.0639999999803	0.639999999932371\\
5.25600000008126	1.28000000018899\\
5.57599999999002	1.91999999997906\\
6.02399999991394	2.55999999987511\\
6.60000000007494	3.2000000000725\\
7.30400000024747	3.84000000020616\\
8.13599999956568	4.47999999969155\\
9.09600000019418	5.12000000012931\\
10.1840000001794	5.76000000010085\\
11.4000000000015	6.40000000001336\\
12.7439999997259	7.03999999987209\\
14.2159999992036	7.6799999996724\\
15.8160000002208	8.32000000009197\\
17.5439999992886	8.95999999975233\\
19.3999999995795	9.59999999985357\\
21.3840000006348	10.2400000001986\\
23.496000000272	10.8800000000756\\
25.7360000007211	11.5200000002027\\
28.1039999989407	12.1599999997216\\
30.6000000010463	12.8000000002618\\
33.2239999997691	13.4399999999485\\
35.9759999999696	14.0799999999816\\
38.8559999983081	14.7199999996277\\
41.8640000007726	15.3600000001575\\
45.0000000016735	16.0000000003356\\
48.2639999996536	16.6399999999347\\
51.6559999984784	17.2799999997164\\
55.1759999972346	17.9199999995063\\
58.8240000043479	18.560000000749\\
62.6000000028944	19.2000000004849\\
66.5040000015797	19.8400000002486\\
67.5040000000263	20\\
150             	20\\
150               	0\\
% 64.535999994594	19.5199999991127\\
% 60.6959999973296	18.8799999995496\\
% 56.9839999989927	18.2399999998177\\
% 53.4000000029409	17.6000000005349\\
% 49.9440000000468	16.9600000000096\\
% 46.6159999994707	16.3199999998989\\
% 43.4159999998265	15.6799999999627\\
% 40.3440000007577	15.0400000001545\\
% 37.4000000022352	14.4000000004909\\
% 34.5839999974705	13.7599999994074\\
% 31.8960000000043	13.1200000000047\\
% 29.3360000000284	12.4800000000041\\
% 26.9039999998713	11.8399999999703\\
% 24.6000000002532	11.2000000000701\\
% 22.4239999988931	10.559999999664\\
% 20.3760000005932	9.92000000018705\\
% 18.4559999995872	9.27999999985643\\
% 16.6640000001571	8.6400000000677\\
% 15.0000000004875	8.00000000019703\\
% 13.4640000010113	7.36000000044165\\
% 12.0559999994111	6.71999999972583\\
% 10.7760000003327	6.08000000018474\\
% 9.62399999958143	5.4399999997593\\
% 8.60000000035507	4.80000000023358\\
% 7.70400000010704	4.1600000000907\\
% 6.93599999993239	3.51999999993255\\
% 6.2960000001367	2.88000000013732\\
% 5.78399999996327	2.23999999993953\\
% 5.39999999996871	1.59999999994139\\
% 5.14399999995294	0.959999999818292\\
}--cycle;
\end{axis}
\end{tikzpicture}%

%% file: root.bbl
% Generated by IEEEtran.bst, version: 1.14 (2015/08/26)
\begin{thebibliography}{10}
\providecommand{\url}[1]{#1}
\csname url@samestyle\endcsname
\providecommand{\newblock}{\relax}
\providecommand{\bibinfo}[2]{#2}
\providecommand{\BIBentrySTDinterwordspacing}{\spaceskip=0pt\relax}
\providecommand{\BIBentryALTinterwordstretchfactor}{4}
\providecommand{\BIBentryALTinterwordspacing}{\spaceskip=\fontdimen2\font plus
\BIBentryALTinterwordstretchfactor\fontdimen3\font minus
  \fontdimen4\font\relax}
\providecommand{\BIBforeignlanguage}[2]{{%
\expandafter\ifx\csname l@#1\endcsname\relax
\typeout{** WARNING: IEEEtran.bst: No hyphenation pattern has been}%
\typeout{** loaded for the language `#1'. Using the pattern for}%
\typeout{** the default language instead.}%
\else
\language=\csname l@#1\endcsname
\fi
#2}}
\providecommand{\BIBdecl}{\relax}
\BIBdecl

\bibitem{guanetti2018control}
J.~Guanetti, Y.~Kim, and F.~Borrelli, ``{Control of connected and automated
  vehicles: State of the art and future challenges},'' \emph{Annual Reviews in
  Control}, vol.~45, pp. 18--40, 2018.

\bibitem{cadillacV2Varticle}
``{V2V Safety Technology Now Standard on Cadillac CTS Sedans},''
  \url{https://media.cadillac.com/media/us/en/cadillac/news.detail.html/content/Pages/news/us/en/2017/mar/0309-v2v.html},
  accessed: 2020-04-01.

\bibitem{uhlemann2016platooning}
E.~Uhlemann, ``{Platooning: connected vehicles for safety and efficiency
  [Connected Vehicles]},'' \emph{IEEE Vehicular Technology Magazine}, vol.~11,
  no.~3, pp. 13--18, 2016.

\bibitem{V2Varticle}
``{Red light, green light - no light: Tomorrow's communicative cars could take
  turns at intersections},''
  \url{{https://doi.org/10.1109/MSPEC.2018.8482420}}, accessed: 2018-10-04.

\bibitem{shladover2012impacts}
S.~E. Shladover, D.~Su, and X.-Y. Lu, ``{Impacts of cooperative adaptive cruise
  control on freeway traffic flow},'' \emph{Transportation Research Record},
  vol. 2324, no.~1, pp. 63--70, 2012.

\bibitem{lioris2017platoons}
J.~Lioris, R.~Pedarsani, F.~Y. Tascikaraoglu, and P.~Varaiya, ``{Platoons of
  connected vehicles can double throughput in urban roads},''
  \emph{Transportation Research Part C: Emerging Technologies}, vol.~77, pp.
  292--305, 2017.

\bibitem{bonnet2000fuel}
C.~Bonnet and H.~Fritz, ``{Fuel consumption reduction in a platoon:
  Experimental results with two electronically coupled trucks at close
  spacing},'' SAE Technical Paper, Tech. Rep., 2000.

\bibitem{shladover2007path}
S.~E. Shladover, ``{PATH at 20 - History and major milestones},'' \emph{IEEE
  Transactions on intelligent transportation systems}, vol.~8, no.~4, pp.
  584--592, 2007.

\bibitem{alam2015heavy}
A.~Alam, B.~Besselink, V.~Turri, J.~Martensson, and K.~H. Johansson,
  ``{Heavy-duty vehicle platooning for sustainable freight transportation: A
  cooperative method to enhance safety and efficiency},'' \emph{IEEE Control
  Systems Magazine}, vol.~35, no.~6, pp. 34--56, 2015.

\bibitem{ploeg2011design}
J.~Ploeg, B.~T. Scheepers, E.~Van~Nunen, N.~Van~de Wouw, and H.~Nijmeijer,
  ``{Design and experimental evaluation of cooperative adaptive cruise
  control},'' in \emph{2011 14th International IEEE Conference on Intelligent
  Transportation Systems (ITSC)}.\hskip 1em plus 0.5em minus 0.4em\relax IEEE,
  2011, pp. 260--265.

\bibitem{milanes2013cooperative}
V.~Milan{\'e}s, S.~E. Shladover, J.~Spring, C.~Nowakowski, H.~Kawazoe, and
  M.~Nakamura, ``{Cooperative adaptive cruise control in real traffic
  situations},'' \emph{IEEE Transactions on Intelligent Transportation
  Systems}, vol.~15, no.~1, pp. 296--305, 2013.

\bibitem{swaroop1997string}
D.~Swaroop, ``{String stability of interconnected systems: An application to
  platooning in automated highway systems},'' 1997.

\bibitem{ploeg2012introduction}
J.~Ploeg, S.~Shladover, H.~Nijmeijer, and N.~van~de Wouw, ``{Introduction to
  the special issue on the 2011 grand cooperative driving challenge},''
  \emph{IEEE Transactions on Intelligent Transportation Systems}, vol.~13,
  no.~3, pp. 989--993, 2012.

\bibitem{van2012cooperative}
E.~van Nunen, M.~R. Kwakkernaat, J.~Ploeg, and B.~D. Netten, ``{Cooperative
  competition for future mobility},'' \emph{IEEE Transactions on Intelligent
  Transportation Systems}, vol.~13, no.~3, pp. 1018--1025, 2012.

\bibitem{smith2019balancing}
S.~W. Smith, Y.~Kim, J.~Guanetti, A.~A. Kurzhanskiy, M.~Arcak, and F.~Borrelli,
  ``{Balancing Safety and Traffic Throughput in Cooperative Vehicle
  Platooning},'' in \emph{2019 18th European Control Conference (ECC)}.\hskip
  1em plus 0.5em minus 0.4em\relax IEEE, 2019, pp. 2197--2202.

\bibitem{hsu1991design}
A.~Hsu, F.~Eskafi, S.~Sachs, and P.~Varaiya, ``{Design of platoon maneuver
  protocols for IVHS},'' 1991.

\bibitem{li1997ahs}
P.~Li, L.~Alvarez, and R.~Horowitz, ``{AHS safe control laws for platoon
  leaders},'' \emph{IEEE Transactions on Control Systems Technology}, vol.~5,
  no.~6, pp. 614--628, 1997.

\bibitem{van2015d3}
J.~van~de Sluis, O.~Baijer, L.~Chen, H.~Bengtsson, L.~Garcia-Sol, and
  P.~Balaguer, ``{D3. 2 Proposal for extended message set for supervised
  automated driving},'' \emph{Brussels, Belgium, Tech. Rep}, vol. 4092490,
  2015.

\bibitem{ploeg2018guest}
J.~Ploeg, C.~Englund, H.~Nijmeijer, E.~Semsar-Kazerooni, S.~E. Shladover,
  A.~Voronov, and N.~Van~de Wouw, ``{Guest editorial introduction to the
  special issue on the 2016 grand cooperative driving challenge},'' \emph{IEEE
  Transactions on Intelligent Transportation Systems}, vol.~19, no.~4, pp.
  1208--1212, 2018.

\bibitem{dolk2017event}
V.~S. Dolk, J.~Ploeg, and W.~M.~H. Heemels, ``{Event-triggered control for
  string-stable vehicle platooning},'' \emph{IEEE Transactions on Intelligent
  Transportation Systems}, vol.~18, no.~12, pp. 3486--3500, 2017.

\bibitem{schindler2020maven}
J.~Schindler, R.~Blokpoel, M.~Rondinone, T.~Walter, O.~Pribyl, H.~Saul,
  A.~Leich, D.~Wesemeyer, and R.~Dariani, ``Maven deliverable 6.4: Integration
  final report,'' 2020.

\bibitem{guzzella2007vehicle}
L.~Guzzella, A.~Sciarretta \emph{et~al.}, \emph{{Vehicle propulsion
  systems}}.\hskip 1em plus 0.5em minus 0.4em\relax Springer, 2007, vol.~1.

\bibitem{turri2015fuel}
V.~Turri, ``Fuel-efficient and safe heavy-duty vehicle platooning through
  look-ahead control,'' Ph.D. dissertation, KTH Royal Institute of Technology,
  2015.

\bibitem{shalev2017formal}
S.~Shalev-Shwartz, S.~Shammah, and A.~Shashua, ``On a formal model of safe and
  scalable self-driving cars,'' \emph{arXiv preprint arXiv:1708.06374}, 2017.

\bibitem{mattingley2012cvxgen}
J.~Mattingley and S.~Boyd, ``{CVXGEN: A code generator for embedded convex
  optimization},'' \emph{Optimization and Engineering}, vol.~13, no.~1, pp.
  1--27, 2012.

\bibitem{lefevre2015learning}
S.~Lef{\`e}vre, A.~Carvalho, and F.~Borrelli, ``{A learning-based framework for
  velocity control in autonomous driving},'' \emph{IEEE Transactions on
  Automation Science and Engineering}, vol.~13, no.~1, pp. 32--42, 2015.

\bibitem{CVXGENinfeasibility}
``{CVXGEN: Code Generation for Convex Optimization - Handling Infeasibility},''
  \url{https://cvxgen.com/docs/infeasibility.html}, accessed: 2020-04-13.

\end{thebibliography}
